\documentclass{aastex7}
\usepackage{amsmath}
\usepackage{amssymb}
\usepackage{color}
\usepackage[dvipsnames]{xcolor}
\usepackage{fancyvrb}
\usepackage{bm}
\usepackage{physics}
\usepackage{pifont} 

\newcommand{\etal}{\emph{et al.}}

\VerbatimFootnotes

\begin{document}

\title{Matched Filtering for the Canadian Hydrogen Observatory and Radio-Transient Detector Galaxy Search}

\newcommand{\perimeter}{Perimeter Institute for Theoretical Physics, 31 Caroline Street N, Waterloo, ON N25 2YL, Canada}
\newcommand{\wca}{Waterloo Center for Astrophysics, University of Waterloo, Waterloo, ON N2L 3G1, Canada}

\author[orcid=0009-0002-1199-8876]{Hans S. Hopkins}
\affiliation{\perimeter}
\affiliation{\wca}
\email{hshopkins@uwaterloo.ca}
\author[orcid=0000-0002-1172-0754]{Dustin Lang}
\affiliation{\perimeter}
\affiliation{\wca}
\email{dlang@perimeterinstitute.ca}
\author[orcid=0000-0002-2088-3125]{Kendrick Smith}
\affiliation{\perimeter}
\email{kmsmith@perimeterinstitute.ca}
\author[0000-0002-0956-7949]{Kristine Spekkens}
\email{kristine.spekkens@queensu.ca}
\affil{Department of Physics, Engineering Physics and Astronomy, Queen's University, Kingston, ON K7L 3N6, Canada}
\author[orcid=0000-0002-0190-2271]{Simon Foreman}
\email{}
\affiliation{Department of Physics, Arizona State University, 550 E.\ Tyler Mall, Tempe, AZ 85287, USA}
\author[0000-0001-7505-5223]{Akanksha Bij}
\email{a.bij@queensu.ca}
\affiliation{Department of Physics, Engineering Physics and Astronomy, Queen's University, Kingston, ON K7L 3N6, Canada}

\begin{abstract}

We present the spatial part of the point source signal extraction strategy for the upcoming CHORD galaxy survey. CHORD, the Canadian Hydrogen Observatory and Radio-transient Detector, is an under-construction drift-scanning compact interferometric radio telescope.  CHORD comprises 512 six meter dishes and observes in the 300 to 1500 MHz frequency range. One of its science goals is producing a catalogue of galaxies detected by the neutral hydrogen (HI) 21 cm emission line. CHORD's highly redundant dish layout creates the problem of spatial aliasing, the effect where the same signal could be feasibly produced from sources at multiple locations on the sky. The search will be done with a matched filter in the visibility plane. This paper presents the search strategy and a prediction tool that can quickly estimate the matched filter response at a given sky position, allowing a prediction of alias locations and severity. This tool confirms that although aliases are impossible to distinguish in a single snapshot, they become possible to distinguish when combining data over a period of time. It predicts that aliases will be harder to distinguish for observations closer to the celestial equator, but that scanning with offset adjacent strips can remove this degeneracy. It predicts that the optimal strategy for a single offset to disambiguate aliases is to re-point the array in declination by about two degrees. A future paper will combine these findings with realistic noise estimates and galaxy population statistics to make forecasts of the population of galaxies that CHORD will detect.

\end{abstract}

\keywords{\uat{Galaxies}{573} --- \uat{Interferometry}{808} --- \uat{Radio astronomy}{1338}}

\section{Introduction}

Large untargeted galaxy surveys are useful for understanding galaxy formation and evolution physics, particularly in the low mass regime. They allow for the measurement of the HI mass function (e.g. \cite{2018_HIMF}, \cite{MA2025}), whose low end constrains the formation rate and survival of low mass objects. They can also be used to determine matter distribution in the local universe (e.g. \cite{Franco_2025}). All-sky HI surveys can be cross-matched with optical surveys in order to guess optical counterparts for the detected galaxies. This allows the study of the relationships between HI gas presence and optical properties like stellar mass and colour. Comparing HI mass and stellar mass gives the HI deficiency of galaxies, which is caused by ram pressure stripping, tidal stripping, and strangulation (e.g. \cite{Lin_2023}). $\Lambda$CDM predicts the existence of dark matter halos with gas but no star formation. There have been efforts to find these objects using untargeted HI surveys (e.g. \cite{du2024opticallydarkgalaxiesdecals}). Untargeted HI survey data was used to find a relationship between dark matter halo mass and HI mass, which can be compared against simulations (e.g. \cite{Obuljen_2019}).

Before 2000, multiple HI wide-field galaxy search surveys were carried out to match the successful optical all-sky surveys, leveraging developments in instrumentation. Some examples are FIRST \citep{FIRST} and NVSS \citep{NVSS} conducted with the Very Large Array, and HIPASS conducted using the Parkes Radio Telescope \citep{HIPASS}. Later efforts increased the survey sensitivity and resolution significantly. Beginning with precursor observations in 2005, a large galaxy search was conducted using the 300 m Arecibo telescope. This survey, called the Arecibo Legacy Fast ALFA (ALFALFA) survey \citep{alfalfafull}, was successful and became the basis of statistical HI galaxy properties analysis \citep{Giovanelli2015, HItoStellarMass}. 

In the 2010 decade, technological advancements in GPUs, telecommunications, and construction materials allowed for more ambitious radio observatories with hopes of carrying out second generation all-sky HI galaxy searches. In 2023, the FAST All-Sky Survey \citep{fashi} detected ten thousand more galaxies than ALFALFA. The WALLABY survey \citep{WALLABY} that is currently underway using the Australian Square Kilometre Array Pathfinder \citep{WallabyPilotPhase2} is projected to detect 100,000 \citep{Deg} galaxies in its the southern hemisphere. Leveraging the expertise and innovations from its predecessor Canadian Hydrogen Intensity Mapping Experiment \citep{chimeOverview}, as well as the aforementioned technological advancements, the Canadian Hydrogen Observatory and Radio-transient Detector (CHORD) \citep{CHORDMainPaper} is also primed to contribute a large HI galaxy catalogue. \citet{CHORDMainPaper} estimated that CHORD could detect up to 300 times as many galaxies as ALFALFA as an upper bound, and a more detailed prediction will be given in Bij \etal\ (in prep.).

CHORD's many-dish interferometric setup comes at a computational cost. Data from all dishes need to be combined and searched for galaxies. CHORD's regular array allows for redundant baseline averaging to help reduce noise and perform calibration, but this creates an aliasing pattern that complicates the search. The CHORD galaxy search will also incorporate frequency information, but we do not discuss that here; the frequency axis can use similar techniques to those used in ALFALFA \citep{Saintonge2007}. This paper addresses the spatial complications that CHORD will face, explains the spatial steps for producing the catalogue, and quantifies the consequences of CHORD's scan strategy for the resulting galaxy catalogue.

Unlike the facilities typically used for HI surveys (e.g. \cite{alfalfafull}), CHORD is a highly redundant array radio interferometric telescope. It will consist of 512  six-meter dishes in a 22$\times$24 rectangular array\footnote{The 22$\times$24 grid is not fully populated due to physical obstructions at the site.}. It will record 6144 frequency channels evenly spaced between 300 and 1500 MHz. This corresponds to 41 km/s velocity resolution at the 21 cm rest frequency and 51 km/s velocity resolution at redshift 0.1. CHORD can optionally use a technique called ``up-channelization" to improve the velocity resolution by an integer factor likely less than or equal to 32 (similar to section 3.2 of \cite{CHIMEFRBMain}). CHORD is suited for a galaxy search because of its sensitivity to radio sources and its drift-scan observation strategy. Over the course of a night, CHORD's pointing relative to the ground will stay fixed, and it will use the Earth's rotation to observe a strip of the sky. Data will be taken continually for each dish, processed, and then the data product will be saved to disk on roughly 10-sec timescales. These integrations will be combined across multiple nights and searched to deduce the location and intensity of galaxies. This paper describes this process, which results in deep effective integration time over a many square degree right ascension strip. Existing single-dish telescopes in this frequency range observe only a handful of pixels at a time.  CHORD, by effectively observing hundreds of pixels simultaneously, observes each patch of sky for more time per day, thereby achieving much greater sensitivity. CHORD's dishes can be re-pointed in declination. One of the goals of this paper is to investigate the advantage that re-pointing CHORD slightly would have on the search for galaxies.

CHORD's high redundancy also produces a substantial spatial aliasing problem, which necessitates care when determining the locations of galaxies. ``Spatial aliasing" is the effect caused by the fact that a source in any of multiple specific relative positions in the sky would give the telescope the same signal. If one naively attempted to map out the sky using interferometric data, a single source would produce multiple bright spots in the resulting map, called aliases. In interferometry, the location of a source on the sky is determined by comparing the phases of its emission received by two dishes. For any two dishes, there are multiple angles on the sky which would produce the same phase difference, causing interferometric spatial aliasing. This is usually mitigated by combining data from pairs of dishes whose vector displacements (``baselines") have different directions, which would cause them to disagree on the alias positions. In CHORD's case, because of its repeated rectangular grid, the baselines are all integer linear combinations of the shortest baseline of the two dimensions. This means that all the dish pairs agree on where the alias positions are. This paper will explain the strategy to handle spatial aliasing and the problems that it poses.

\section{Methods}

The strategy for the galaxy search is different than the search strategy for other CHORD science cases. The position of aliases depends on frequency. If the target signal is extended in frequency, as is the case for Fast Radio Bursts and pulsars, then this fact can be used to identify aliases. Galaxies are relatively narrow in frequency space; the velocity range of a small dispersion-supported galaxy is on the order of 10 km/s, which is narrower than a CHORD frequency channel. More typical galaxies (for example ones selected to measure the Tully-Fisher relation from the ALFALFA catalogue \cite{tully-fisher-alfalfa}) have a velocity width on the order of 100 km/s, spanning multiple frequency channels, but they are still too spectrally narrow to detect aliases using their frequency dependence alone. The search strategy for galaxies must also differ from CHORD's hydrogen intensity mapping pipeline which handles large spatial scale information. Galaxies will appear pointlike instead. The FWHM of CHORD's synthesized beam is 7.7 arcminutes at a redshift of 0.1 in the east-west direction. A Milky Way sized galaxy at that distance would be only 14 arcseconds across. The galaxy search is therefore designed to pick out spectrally narrow, spatially pointlike sources. It will take advantage of the expected galaxy spectrum, but this paper will specifically only explain the spatial aspect of the search which arises from the driftscan observing method. For this paper, the frequency will be fixed for all scenarios, as if each galaxy only occupies a single spectral channel. This paper will also treat each galaxy as isolated within a pixel, which is a valid assumption for low redshift in low galactic density density environments.

Table \ref{tab:fidvalues} summarizes the fiducial values that are used for computing the examples unless otherwise specified.
\begin{center}
    \begin{table}
    \begin{tabular}{ccccc}
        \hline
         $m_1 \times m_2$ & Wavelength & Primary beam FWHM & Synthesized beam FWHM & Declination range \\ \hline
         24 $\times$ 22 & 0.21 m & 2.0 deg & 3.1' $\times$ 7.0' & 20-80 deg\\
         \hline
    \end{tabular}
    \caption{Fiducial values that are used in the plots unless otherwise specified.}
    \label{tab:fidvalues}
    \end{table}
\end{center}

\subsection{Visibilities and Derived Quantities}

When CHORD is observing, it continuously measures voltage readings from each antenna, frequency-channelizes them, correlates these signals between antennas and averages over a time scale of tens of seconds before saving the results to disk. This data product is called a visibility matrix or  visibilities, and is an array indexed by dish pair, frequency, and polarization. For this paper, we will be ignoring frequency and polarization. It is complex valued to record the amplitude and phase of the correlation between dishes. Phase offsets between the electric fields measured by the antennas correspond to the sky directions that the wave originates from. In an idealized scenario with no noise, the visibilities that CHORD would save for one integration period are
\begin{equation}
    V_{jk} = \int_{0}^\pi \int_0^{2\pi} I(\phi, \theta) \ B^2(\abs{\boldsymbol n(\phi,\theta) - \boldsymbol n_p}) \ \exp\left(\frac{2\pi i}{\lambda} (\boldsymbol x_j - \boldsymbol x_k)\cdot \boldsymbol n(\phi,\theta) \right) \ \text d\phi \text d\theta \quad,
\label{eqn:visibilities}
\end{equation}
where $\boldsymbol x_j$ is the position of the $j$th dish in the array represented as a 3-vector, $\lambda$ is the observing wavelength, $\boldsymbol n(\phi,\theta)$ is a sky unit vector, and $I(\phi,\theta)$ is the intensity from the sky \citep{Liu_2020}. A location on the sky far away from where the dishes are pointing will contribute diminished power to the visibilities depending on the dish shape. The \textit{primary power beam} response of CHORD is given by $B^2$, a function of the angular difference between the location on the sky and the direction the dishes are pointing $\boldsymbol n_p$. This power falloff effect determines the width of the strip of sky that CHORD can observe for a single declination pointing while drift-scanning. For this paper, the primary beam will be modeled as an Airy beam, which is appropriate for a circular dish:
\begin{equation}\label{eqn:primarybeam}
    B^2(\alpha) = \left(\frac{2 J_1(\tilde \alpha)}{\tilde \alpha}\right)^2; \quad \tilde \alpha = \frac{\pi D \sin(\alpha)}{\lambda} \quad ,
\end{equation}
where $J_1$ is the Bessel function of the first kind and $D$ is the dish diameter (6 m for CHORD).

In the case when the sky consists of a single point source at $\boldsymbol n_s$, $I(\phi,\theta)$ is a delta function. The integral evaluates to
\begin{equation}\label{eqn:singlesourcevisibilities}
    V_{jk} = I_s B^2(\abs{\boldsymbol n_s - \boldsymbol n_p})\exp\left(\frac{2\pi i}{\lambda} (\boldsymbol x_j - \boldsymbol x_k)\cdot \boldsymbol n_s \right) = I_s t(\boldsymbol n_s) \quad,
\end{equation}
where $I_s$ is the intensity of source $s$. Equation \ref{eqn:singlesourcevisibilities} serves as a \emph{template} for the visibility response of a point source object $t(\boldsymbol n_s)$ which can be used in a search algorithm. This template is a vector indexed by the $jk$ dish pair. The rest of this paper will discuss quantities that are derived from this template acting with some visibility data $d$.

During each integration period, CHORD takes billions of voltage measurements. The averaged saved visibilities are therefore in the central limit theorem regime, so each element of the visibility array is normally distributed. The covariance of the visibilities will be represented by the matrix $N$. For this paper, we will assume that the noise is the same for every dish, that there is no correlation between the noise produced by any dishes, and that the noise does not change over time. This assumption is invalid in frequency regimes where radio frequency interference (RFI) is important, but it is expected that RFI can be excised at low redshifts. The fact that visibilities are normally distributed lets us use a Gaussian statistics framework to define quantities that can be used in a search algorithm or to process the visibilities into sky maps.

There are two quantities that are useful for quantifying the sky map that these visibilities produce. We define the Likelihood Change Map (LCM), an intermediate quantity that can be used in the computation of the others. It is defined as
\begin{equation}\label{eqn:lcm}
    \text{LCM}(\boldsymbol n;d) = \text{re}\left(t^\dagger(\boldsymbol n) N^{-1} d\right) \quad,
\end{equation}
where the dagger symbol indicates conjugate transpose. Because of the form of the template, this matrix multiplication is a discrete Fourier transform. Taking the real part of this multiplication indexed by one-way baselines is equivalent to taking the Fourier transform indexed by two-way baselines and dividing by 2. The LCM is convenient because it is linear in each visibility. Different LCMs can be added together without changing the normalization. For example, LCMs captured on different days can be summed together when the same sky pixelization is used. The physical interpretation of this map is that it represents the increase in the log-likelihood (assuming Gaussian noise) of observing the visibilities $d$ per unit brightness of a source added in a particular location $\boldsymbol n$. It therefore has units of inverse specific intensity. It can be used to visualize locations on the sky where a source has probably produced a strong response with the instrument. The Likelihood Change Map is an example of a ``dirty map" with a natural weighting (Ch.~7 in \citealt{Perley}). ``Dirty map" is an ambiguous term in radio astronomy referring to a sufficient statistic of the true sky produced by a simple linear operation on the visibilities. Dirty maps can have different visibility weightings or be normalized to have units of specific intensity instead of inverse specific intensity, for example the quantity $t^\dagger(\boldsymbol{n})d$.

The Maximum Likelihood Map represents the distribution of brightness on the sky which optimizes the likelihood of observing the visibilities ($d$). It is defined as
\begin{equation}
    \text{MLM}(\boldsymbol n;d) = \text{re}\left(\frac{t^\dagger(\boldsymbol n) N^{-1} d }{t^{\dagger}(\boldsymbol n) N^{-1} t(\boldsymbol n)}\right) \quad .
\end{equation}
The advantage of this map is that it has units of specific intensity and more closely matches the expectation of what a sky map should represent. The disadvantage is that the normalization changes CHORD is pointed in a different direction, making addition of maps more complicated, and the MLM diverges rapidly in regions where the instrument does not have a strong response. After each galaxy is found, one can compute the maximum likelihood map at the position of the galaxy to get the best estimate of the galaxy's brightness.

Neither the likelihood change map nor the maximum likelihood map are optimal candidates for the quantity that we should use to find galaxies. Both of them will have peaks at the locations of real sources, but the heights of those peaks are difficult to interpret. Because of aliasing, we want to be able to reject peaks based on their values. The likelihood change map suppresses pixels by a factor of the primary beam, so a peak that is a primary beam away from CHORD's pointing center could be confused for noise or an alias because of its low value. The maximum likelihood map has the opposite problem because it amplifies values away from the telescope's pointing center too strongly. The best quantity to use is the ``matched filter" because it has the correct normalization for the problem. The matched filter is a mathematical tool that is occasionally used in astronomy for searching for a well-understood template in noisy data (e.g. \cite{BOSS}, \cite{redmapper}, \cite{Melin}, \cite{Willman}, \cite{Doyle_2000}, \cite{ligo}). 

The matched filter is the optimal linear filter in the sense that it maximizes detection signal-to-noise ratio (SNR). Given a dataset $d$ with Gaussian noise and a template $t(n)$ that depends on template parameter set $n$, the matched filter represents the probability that the data are a scaled template plus noise, in units of signal to noise ratio. The data vector $d$ scales with signal and the covariance matrix $N$ scales as the square of the noise. The matched filter $M(n;d)$ is defined as
\begin{equation}\label{eqn:matchedfilter}
    M(n; d) = \text{re}\left(\frac{t(n)^\dag N^{-1} d}{\sqrt{ t(n)^\dag N^{-1} t(n)}}\right) \quad .
\end{equation}

The proof that this definition fulfills the requirements for a matched filter can be found in appendix A of \cite{hopkinsthesis}. The units of the matched filter make its interpretation of a detection easier than the other two quantities. Any high value of $M$ indicates that the template at $n$ ``matches" well with the data. The numerator checks the closeness between the template and the data. The denominator normalizes the result to be in signal-to-noise units. In practice, we will compute $M(n;d)$ for a large set of $n$ templates covering the parameter space of the template model that we wish to search. Peaks in $M(n;d)$ will be found, and the corresponding best $n$ values will be returned.

In this case, the template parameter $n$ is the proposed location on the sky for a galaxy $\boldsymbol n$. The vector $d$ contains the observed visibilities, and the vector $t$ contains the theoretical response visibilities of a single point source, given by equation \ref{eqn:singlesourcevisibilities}. $M(\boldsymbol n;d)$ will be computed for a sky map where each pixel of the map provides the $\boldsymbol n$. Unlike the LCM and MLM, if a peak is detected far away from where CHORD is pointing, it can still be interpreted as a likely location for a galaxy. However, a high value of a pixel of $M(\boldsymbol n;d)$ does not necessarily indicate that a galaxy is present in that pixel. Multiple pixels can have high values caused by the same source. If the pixelization of the sky is oversampled enough, then all pixels nearby to the position of a galaxy will have high values, as well as pixels in alias locations. The probabilistic interpretation is that the matched filter values for these other pixels are being drawn from a noisy distribution which is $R(\boldsymbol{n}', \boldsymbol n)$ correlated with the matched filter value of the true location. It is important to know which pixels have large values caused a single source and not count those multiply.

We define the correlation coefficient $R(\boldsymbol n', \boldsymbol n;d)$ between two pixels $\boldsymbol n'$ and $\boldsymbol n$ on the sky as
\begin{equation}\label{eqn:correlationcoeff}
R(\boldsymbol n', \boldsymbol n)  = 
\frac{t(\boldsymbol n)^\dag N^{-1} t(\boldsymbol n')}{\sqrt{t(\boldsymbol n)^\dag N^{-1} t(\boldsymbol n) } 
\sqrt{t(\boldsymbol n')^\dag N^{-1} t(\boldsymbol n')}} \quad .
\end{equation}

This will range in value between 0 for empty sky and 1 at the location of the true source. This will tell us not only the location of aliases, but the value indicates how much a pixel at an alias location would correlate to a pixel at the true location. If an alias location is only weakly correlated to the true location of the source, then the corresponding peak in the matched filter will be lower, allowing an algorithm to distinguish which peak is real. On the other hand, with the addition of random noise, there is the danger that a nearly 100\% correlated alias will randomly get boosted above the level of the true source and therefore be confused with the true source. The majority of this paper will be focused on investigating the correlation coefficient map for various cases to check when there are dangerously high aliases.

\subsection{Method Consequences and Solutions}

This paper focuses on investigating aliases because they can lead to three problems when producing a galaxy catalogue with CHORD. First, after identifying a galaxy candidate, an algorithm might infer that the galaxy is at an aliased position rather than the true source position because of an unlucky noise fluctuation. This would result in the corresponding catalogue entry reporting the incorrect localization and total power. Second, a faint source near the location of an aliased bright source might be drowned out by the contribution from the bright source. Thirdly, there might be false positive detections because of aliases from sources outside of the scanning area. CHORD is a drift-scan telescope, so it will observe strips of the sky at a time. If a source is outside of the strip but close enough that its aliases are inside of the strip, those aliases might be identified as real sources. Addressing all of these issues will require knowledge on the locations of aliases and their severity, both of which will be visualized by plotting the correlation coefficient under various conditions.

The CHORD operators will have control over its declination as well as how often it is re-pointed. We will show that the correlation coefficients of aliases depend on the chosen declination, and that they can be significantly reduced with an offset-adjacent-strip scan strategy.

The remainder of this paper will be divided into three sections. For each of them, we will derive an explicit equation for the correlation coefficient $R(\boldsymbol n, \boldsymbol n')$ for increasingly general situations. First, we will present the equation for the case of an instantaneous capture of the visibilities. Second, we will consider the case where visibilities measured of an entire sidereal day are treated jointly by the matched filter. Third, we will show the effects of adjacent-strip scanning: combining visibilities observed at different CHORD pointings. Each of these equations will be derived for an instrument that consists of a perfectly rectangular array of dishes with no gaps. The plots will show the results for a 22$\times$24 dish array, which has more dishes than CHORD because CHORD will have gaps in its layout. Some of the dishes being missing is not expected to change the conclusions of the paper qualitatively.

\section{Instantaneous analysis}
\label{sec:instantaneous}

For the instantaneous case, we will simulate running the matched filter on the visibilities of a single point source in the CHORD field of view at one time. The template is given by the visibilities of a single point source located at $\boldsymbol n_s$ (equation \ref{eqn:singlesourcevisibilities}). The overall scaling of the template does not matter, so we drop constant factors:
\begin{equation}\label{eqn:spatialtemplate}
    {t(\boldsymbol n_s)}_{jk} = B\left(\abs{\boldsymbol n_s - \boldsymbol n_p}\right)^2 \exp\left(\frac{2\pi i}{\lambda} (\boldsymbol x_j - \boldsymbol x_k)\cdot \boldsymbol n_s \right) \quad.
\end{equation}

The next step is to compute the correlation coefficient (equation \ref{eqn:correlationcoeff}). We will start with the numerator $t^\dagger N^{-1} t$. It can be shown that since we are assuming that the noise draws for each dish are independent, the expectation value for the correlation of the noise from two distinct dish pairs is 0, so the covariance matrix is diagonal. The assumption that the noise is identical from each dish means that every element of the covariance matrix corresponding to an autocorrelation is the same. We will call this value $\frac{1}{\sigma^2}$. It can be shown \citep{NRAOlectures} that the elements corresponding to cross-correlations are equal to $\frac{2}{\sigma^2}$. If we substitute $t$ and $N$ into the numerator of equation \ref{eqn:correlationcoeff} and simplify, we get the expression
\begin{equation}
    B\left(\abs{\boldsymbol n - \boldsymbol n_p}\right)^2 B\left(\abs{\boldsymbol n_s - \boldsymbol n_p}\right)^2\left(\frac{2}{\sigma^2}\sum_{j= k+1}^{m_1m_2}\sum_{k=1}^{m_1m_2} \exp\left(\frac{2\pi i}{\lambda} (\boldsymbol x_j - \boldsymbol x_k)\cdot (-\boldsymbol n + \boldsymbol n_s) \right) + \frac{m_1m_2}{\sigma^2}\right) \quad ,
\end{equation}
where $m_1$ and $m_2$ are the number of dishes in the two dimensions of the dish grid. The last term comes from the autocorrelations. The double sum is from the cross-correlations. Because swapping $j$ and $k$ does not change the value of the exponential, the sum can absorb the factor of 2 and be rewritten as a double sum that counts each dish pair twice. The autocorrelations term can also be absorbed into the sum as the $j=k$ case. With these simplifications and cancellations from the denominator, the correlation coefficient can be written as
\begin{equation}
    R(\boldsymbol n, \boldsymbol n_s) = \frac{1}{m_1^2 m_2^2} \sum_{j=1}^{m_1m_2} \, \sum_{k=1}^{m_1m_2} \exp\left(\frac{2\pi i}{\lambda} (\boldsymbol x_j - \boldsymbol x_k)\cdot (-\boldsymbol n+\boldsymbol n_s) \right) \quad .
\end{equation}
 This computation would involve summing $m_1^2 m_2^2$ terms, which is time consuming. With the assumptions made about a fully-populated rectangular array, the sums in the correlation coefficient can be separated, and the following algebra fact can be used to simplify this expression into a form that is faster to compute. More detail is given in \cite{hopkinsthesis} section 2.3.
\begin{equation}\label{eqn:usefulfact}
    \sum_{j=1}^{m} \exp(ic(\tfrac{-m-1}{2}+j)) = \frac{\sin(cm/2)}{\sin(c/2)}
\end{equation}

\begin{equation}\label{eqn:instantaneousmainresult}
    R(\boldsymbol n, \boldsymbol n_s) = \frac{1}{m_1^2 m_2^2} \frac{\sin(\pi m_1 L_1/\lambda \ \boldsymbol e_1\cdot (-\boldsymbol n+\boldsymbol n_s))^2}{\sin(\pi L_1/\lambda \ \boldsymbol e_1 \cdot (-\boldsymbol n+\boldsymbol n_s))^2}\frac{\sin(\pi m_2 L_2/\lambda \ \boldsymbol e_2\cdot (-\boldsymbol n+\boldsymbol n_s))^2}{\sin(\pi L_2/\lambda \ \boldsymbol e_2 \cdot (-\boldsymbol n+\boldsymbol n_s))^2}
\end{equation}
where $\lambda$ is the wavelength, and $L_2$ is the east-west separation between neighboring dishes in the grid, and $\boldsymbol e_1$ and $\boldsymbol e_2$ are the two basis vectors of the grid. $L_1$ is the separation between neighboring dishes north-south direction times the cosine of the difference between the pointing declination and CHORD's zenith declination. When the dishes are pointing straight upward, $L_1$ corresponds to the physical spacing between the dishes.  When pointing at different elevations, the effective baseline becomes the projected separation (perpendicular to the dish pointing direction) as a foreshortening effect. This form of $R(\boldsymbol n, \boldsymbol n_s)$ is much quicker to compute than the original; we can now compute it for a large grid of $\boldsymbol n$ directions and display the result as an image.

\begin{figure}[htb!]
\begin{center}
\includegraphics[width=0.5\textwidth]{"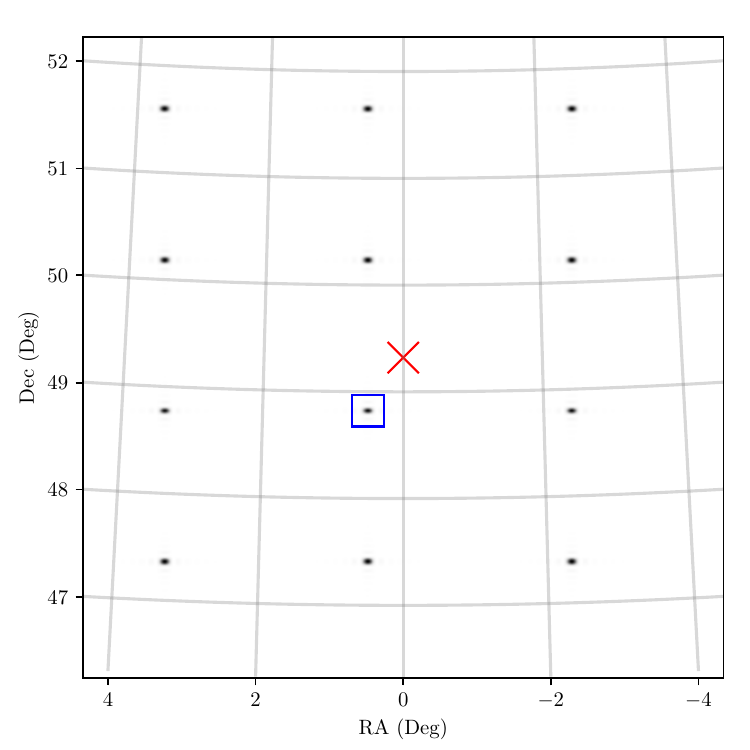"}
\end{center}
\caption{The correlation coefficient $R(\boldsymbol n, \boldsymbol n_s)$ (equation \ref{eqn:instantaneousmainresult}) plotted in equatorial (RA, Dec) sky coordinates. The direction that CHORD is pointing at an instant, $\boldsymbol n_p$, is indicated with the \ding{53}.  In this case, CHORD is pointed at its zenith. The source position $u_s$ is indicated by the box $\Box$. The wavelength is set to 21 cm. A grid of locations yield $R(\boldsymbol n, \boldsymbol n_s) = 1$, which are perfect aliases.}
\label{fig:instantaneous}
\end{figure}

Figure \ref{fig:instantaneous} shows a plot of equation \ref{eqn:instantaneousmainresult} (where each pixel represents a $\boldsymbol n$), covering a few square degrees of sky. The wavelength was chosen to be 21 cm as an example, but other choices of wavelength for the purposes of the galaxy search are close enough to not change the qualitative takeaways. Aliases appear on a rectangular grid in tangent-plane coordinates corresponding to a grid of east-west and north-south vectors projected onto the sky. They are separated by about 1.5 degrees. The alias grid spacing can be derived on paper from equation \ref{eqn:instantaneousmainresult}. Since the sine is squared and $m_\rho$. $\rho$ is a dummy index representing grid dimension 1 or 2. The fraction will evaluate to $m_\rho$ whenever
\begin{equation}
    \frac{L_\rho}{\lambda}\boldsymbol e_\rho\cdot (-\boldsymbol n+\boldsymbol n_s) \in \mathbb{Z} \quad .
\end{equation}
This indicates that the alias separation close to the source is scaled by $\frac{\lambda}{L_\rho}$.

For a more heuristic explanation, we expect the alias correlation coefficients to be 1 because, for an observation at a single time, a redundant-baseline array cannot distinguish brightness coming from those aliased locations from brightness of the true source.  We expect sources farther away from the primary beam centre to produce a dimmer response, but this dimming factor is canceled out by the denominator of the matched filter equation. This is so that potential sources that are farther away from the primary beam are weighted properly. In other words, the matched filter knows that the primary beam will attenuate sources far from CHORD's pointing direction, and it compensates for this to calculate the correct SNR.

It is clear that a search for galaxies using a matched filter on a single visibility matrix snapshot will suffer badly from the existence of perfect aliases. In the next section we will show that adding in information from multiple time samples can break this degeneracy.

\section{Time-integrated analysis}
\label{sec:integrated}

The template defined in equation \ref{eqn:singlesourcevisibilities} describes the instantaneous visibilities that would be measured from a single source. These visibilities would be produced after a 10--30 second integration period, provided there is fringestopping to ensure that the source's movement during this period does not cause any smearing. CHORD's primary beam has a width of about 2.4 $\cos(\text{dec})$ degrees, which corresponds to dozens of 20 second periods, so over the course of a night, several saved visibility matrices will include signal  from the same source as it passes overhead. These can be integrated to provide a higher signal measurement of each source. This time integration has the side effect of making the aliases distinguishable in some cases. This can be seen by representing the integration using the template array framework. A new template can be defined with a ``time step" index alongside the dish indices: $t_{\tau jk} (\boldsymbol n_s)$. This represents the instantaneous visibilities at the $\tau$th time step, up to $n_\tau$.

The passage of time will be symbolically represented by the rotation of the sky by the rotation matrix $\mathcal R(\omega \tau)$. $\omega$ is the rotational velocity of the Earth. The $\tau = 0$ point is arbitrary, but it can be viewed as the beginning of an observing night. $\boldsymbol n$ is replaced by $\mathcal R(\omega\tau)\boldsymbol n$ and $\boldsymbol n_s$ is replaced by $\mathcal R(\omega\tau)\boldsymbol n_s$. With these replacements, the new template is
\begin{equation}\label{eqn:integratedtemplate}
    t(\boldsymbol n, \tau) = B\left(\abs{\mathcal R(\omega \tau)\boldsymbol n - \boldsymbol n_p}\right)^2 \ \exp\left(\frac{2\pi i}{\lambda} (\boldsymbol x_{j} - \boldsymbol x_{k}) \cdot \mathcal R (\omega\tau)\boldsymbol n\right) \quad .
\end{equation}

This template includes information about the increase and decrease of power as the source passes over the primary beam, as well as the path that the source travels on across the sky, which is a line of constant declination and changing hour angle.

Placing this template into the matched filter equation (\ref{eqn:matchedfilter}) reveals that the numerator and denominator do not cancel as much as in the instantaneous case. In particular, the primary beam terms do not cancel:
\begin{multline}\label{eqn:integratedmain}
    R(\boldsymbol n, \boldsymbol n_s) = \frac{1}{n_\tau m_1^2 m_2^2}\frac{1}
      {
      \sqrt{\sum_\tau B\left(\abs{\mathcal R(\omega \tau)\boldsymbol n - \boldsymbol n_p}\right)^4}\sqrt{\sum_\tau B\left(\abs{\mathcal R(\omega \tau)\boldsymbol n_s - \boldsymbol n_p}\right)^4}}\\
   \times \sum_{\tau} 
      B\left(\abs{\mathcal R(\omega \tau)\boldsymbol n - \boldsymbol n_p}\right)^2
      B\left(\abs{\mathcal R(\omega \tau)\boldsymbol n_s - \boldsymbol n_p}\right)^2\\
      \times \frac{\sin(\pi m_1 L_1/\lambda \ \boldsymbol e_1\cdot R(\omega \tau)(-\boldsymbol n+\boldsymbol n_s))^2}{\sin(\pi L_1 / \lambda \ \boldsymbol e_1 \cdot R(\omega \tau)(-\boldsymbol n+\boldsymbol n_s))^2}\frac{\sin(\pi m_2 L_2 / \lambda \ \boldsymbol e_2\cdot R(\omega \tau)(-\boldsymbol n+\boldsymbol n_s))^2}{\sin(\pi L_2 / \lambda \ \boldsymbol e_2 \cdot R(\omega \tau)(-\boldsymbol n+\boldsymbol n_s))^2}.
\end{multline}

For a single time step, there are always completely degenerate aliases. With multiple time steps, there is an opportunity to break these degeneracies. Two effects contribute to this degeneracy breaking: one is based on how the primary beam affects the brightness of the alias as a function of time; the other is based on the location of the alias due to the curvature of the sky. These two effects can be seen in the two different places $\mathcal R(\omega\tau)$ appears in equation \ref{eqn:integratedmain}. These effects are detailed in the following paragraphs.

The first effect is that the templates at the true source location and the alias locations see different parts of the primary beam function. This effect applies to all alias locations. The matched filter expects to see the strongest signal when the source passes closest to the center of the primary beam; when the source is far from the primary beam, the template is suppressed. For aliased source positions, the matched filter expects the signal to peak at the wrong time, so although the individual time steps are scaled copies of the instantaneous templates, their relative amplitudes are mis-matched with the observed signal. This mismatch causes the matched filter to produce a smaller signal for the aliased position relative to the true position in the integrated case.

\begin{figure}[htb!]
\plotone{"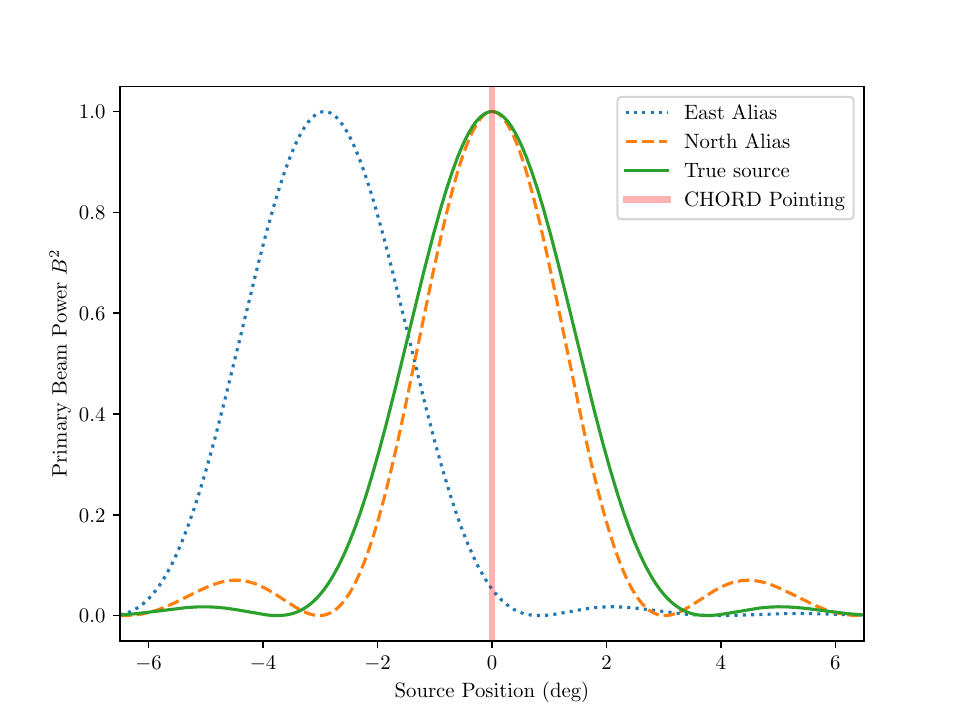"}
\caption{Predicted primary beam power for the true source location and closest alias locations as they move across the meridian. The north alias is about 1.4 degrees away, and the east alias is about 1.9 degrees away. Since the source is moving across the sky at a constant rate, the horizontal axis also corresponds to time. The source is located at the declination CHORD is pointed to.
The matched filter expects the signal to peak as the source transits the meridian, passing closest to the center of the primary beam, as shown in the ``True source'' curve.  For an aliased position to the east (blue dotted curve), the matched filter expects the signal to peak at the wrong time. For an aliased position to the north, the peak occurs at the correct time, but the detailed shape of the peak and its lobes do not match the data perfectly.
}
\label{fig:pbcrossing}
\end{figure}

This effect can be seen in equation \ref{eqn:integratedmain}. Ignoring the sine fraction part, the correlation coefficient looks like the correlation between the alias's primary beam value and the source's primary beam value. It is 1 if $\boldsymbol n = \boldsymbol n_s$, and less than 1 otherwise. Figure \ref{fig:pbcrossing} shows how the correlation would drop below 1 for aliases to the east and to the north.
Aliases in the east-west direction (at different RAs than the true source) 
have their primary-beam peaks at the wrong times.  This significantly reduces their
time-integrated correlation coefficients.
Aliases that are purely in the north-south direction (ie, at the same RA as the true source) have only slightly different primary-beam response curves compared to the true source, so can still have relatively large correlation coefficients.

\begin{figure}[htb!]
\plotone{"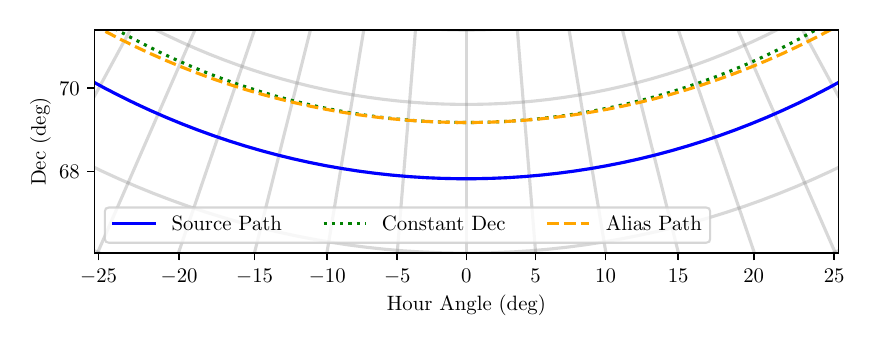"}
\caption{The deviation of an alias's path from a path of constant declination. In this example, CHORD is pointed at a high 70 degree declination, and there is a source at that same declination. The position of the first northern alias is shown with a dashed line.  In this tangent-plane projection, the alias follows a path that is a constant vertical offset from the true source path.  In contrast, a real source would follow a constant-declination path, shown with a dotted line.
These paths diverge slightly at large hour angles.
}
\label{fig:curvature}
\end{figure}

The second effect that helps to reduce aliasing is the difference in the paths followed by aliases and real sources through the sky. Real sources follow paths with constant declination, but aliases do not. Aliases are fixed relative to the position of the real source in tangent-plane coordinates instead. The difference between consecutive instantaneous alias locations and the path that a real source at the same declination of the alias when it passes over CHORD increases with hour angle (HA) away from CHORD and is measurable. When the north alias leaves CHORD's primary beam at an HA of about 2 degrees, its path diverges from the path of a real source by about 3 arcminutes, which is about the synthesized beam FWHM scale of CHORD. This misalignment will decrease the correlation between the matched filter testing a template placed at the alias location and a template that exactly follows the source for every time step. This effect is stronger at the poles and weaker at the equator since it relies on difference of curvature. In the flat sky approximation, this effect is nonexistent. This effect also only applies to aliases which have different declinations than their corresponding true source. The difference in paths is shown in Figure \ref{fig:curvature}. This effect also works in conjunction with the previous effect by adding to the distortion of the primary beam crossing curves that can be seen in Figure \ref{fig:pbcrossing}.

This effect can be seen mathematically by comparing the analysis at the end of section \ref{sec:instantaneous} with equation \ref{eqn:integratedmain}. For any individual term in the sum, it is possible to find a $\boldsymbol n_a$ other than $\boldsymbol n_s$ which produces the same value for the sine fractions. This $\boldsymbol n_a$ will depend on the relative positions of CHORD and the source, so a viable $\boldsymbol n_a$ for one time step will no longer work in other time steps. This causes the matched filter to drop below 1 after summing over multiple time steps.

\subsection{Alias Analysis}\label{sec:aliasanalysis}

\begin{figure}[htb!]
\centering
\includegraphics[height=0.4\textwidth]{"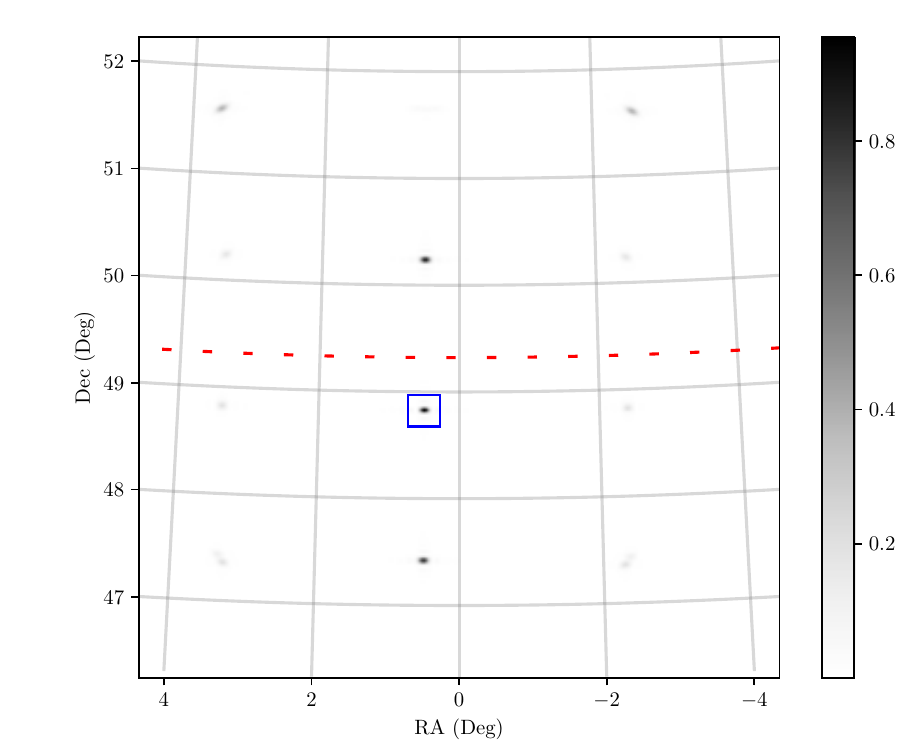"}
\includegraphics[height=0.4\textwidth]{"instantaneous_fig.pdf"}
\caption{\emph{Left:} The correlation coefficients given by equation \ref{eqn:integratedmain} with CHORD pointing straight up and a source passing nearby. 
The dashed line indicates the path of CHORD's pointing direction.
The true source position is indicated by the box $\Box$.
The first aliases to the north and south remain prominent, while the aliases to the east
and west are significantly reduced in amplitude.
\emph{Right:} a copy of Figure \ref{fig:instantaneous}, the instantaneous case, for comparison.}
\label{fig:integratedone}
\end{figure}

Figure \ref{fig:integratedone} shows the time-integrated matched-filter correlation coefficients for a typical case. This illustrates that the perfect aliases in the instantaneous case are no longer perfect when considering multiple time samples. The aliases still form nearly the same grid pattern with the same spacing, but they are suppressed to various degrees. The two effects discussed above can be seen; the east and west aliases are highly suppressed because of the primary beam crossing effect. The north and south aliases are affected less by the primary beam crossing effect, and the curvature effect is less significant. The aliases can also change shape, which can be seen with the two aliases in the upper corners. There is no longer a symmetry that guarantees that they all look identical.

\begin{figure}[htb!]
\plotone{"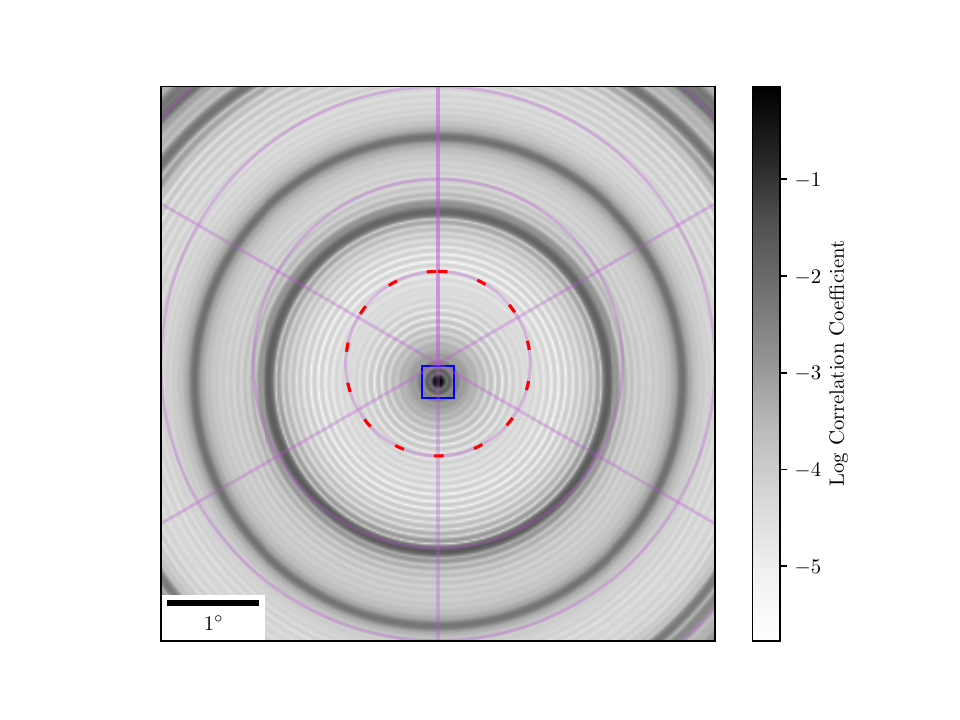"}
\caption{A map of the log correlation coefficients for the case of a source near the north pole. The position of the source is indicated by the box $\Box$. The purple declination grid lines are separated by 1 degree. CHORD's pointing declination at 89 degrees is indicated by the dashed line.  The concentric rings of higher correlation coefficient are aliases that rotate around the true source position at a fixed angular distance over the course of a sidereal day.  This smears out the power of the aliases into rings rather than discrete peaks.}
\label{fig:integratedtwo}
\end{figure}

We are interested in how this picture changes depending on CHORD's pointing declination and the source position. Figure \ref{fig:integratedtwo} shows an exaggerated scenario where CHORD is pointed only one degree away from the north pole, which will not happen in any real observation. Being so close to the north pole makes the effect of curvature much more pronounced. The alias paths deviate significantly from the path of a source in the alias's position, which causes the alias power to be spread out in rings. The alias power being spread out means that the likelihood of any individual alias position being misidentified as a true source is greatly decreased. Pointing more north makes alias disambiguation easier because the curvature effect is stronger. An Airy ring pattern can be seen around the source.

We will now move away from showing maps to focus on the correlation coefficients of the nearest aliases to the source. These aliases are the ones that an algorithm would be most likely to confuse with the true source. This negative outcome is called ``mislocalization" because it would result in the algorithm reporting the incorrect pixel as the location of the source. This section will investigate the severity of nearby aliases depending on the scan strategy of CHORD. 
To produce these plots, we find which $\boldsymbol n$ corresponds to the position of a nearby alias, and then we plot the correlation coefficient of that $\boldsymbol n$ with the true source location $\boldsymbol n_s$.

The following plots explore the correlation coefficient predictions depending on CHORD's pointing. The parameters being tested are CHORD's pointing declination, the relative positions of the source and CHORD, and using multiple pointings. We explore these options by simulating the same base case, CHORD pointing at zenith with a source about half a degree away, and then vary one parameter at a time.

\begin{figure}[htb!]
\plotone{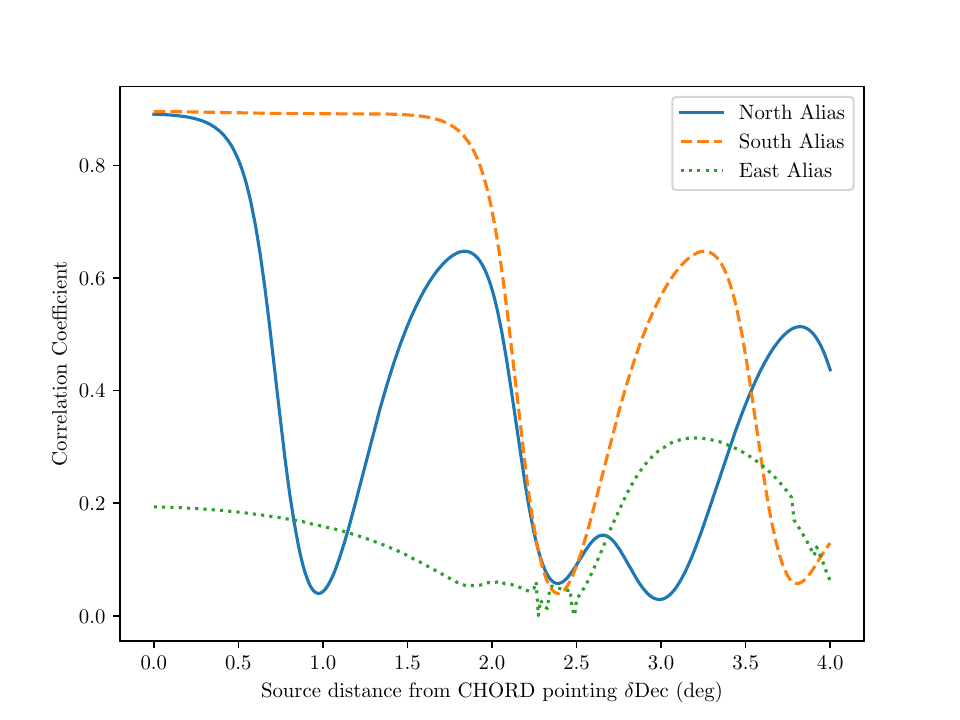}
\caption{The severity of aliasing as a function of the source's position within the primary beam.  In this simulation, CHORD is pointing at the zenith, and we vary the declination of the source.  A $\delta$Dec of zero indicates that the source passes through the center of the primary beam, and a positive $\delta$Dec indicates a source that is north of CHORD's pointing.
CHORD is most sensitive to a strip about 1 degree north and south of the declination of its pointing. A catalogue that is produced from a single strip will have entries with aliasing effects that vary depending on the declination of the galaxies within that strip. Specifically the north and south aliases will vary significantly.
At some values of $\delta$Dec, the east alias becomes bimodal, and its value becomes difficult to track. However, at a correlation coefficient value of around 0.1, an alias has a negligible chance of causing mislocalization.}
\label{fig:deltatheta}
\end{figure}

The first configuration tested is fixing CHORD's pointing at its zenith, and then moving the source away from it in the declination direction by an angular distance $\delta$Dec. The results are plotted in Figure \ref{fig:deltatheta}. As a general trend, alias disambiguation becomes easier the farther away it is because the primary beam crossing effect is stronger. The curves are not monotonic because of the non-monotonic shape of the primary beam. Alias disambiguation being easier if the source is farther away does not mean that we should aim CHORD such that the targets are on the very edge of the strip. CHORD's sensitivity drops off near the edge of the strip, so targets should still be placed in the center of the strip in order to optimize detection probability. The exception would be if the target's detection is guaranteed and the only priority is minimizing alias mislocalization.

\begin{figure}[htb!]
\plotone{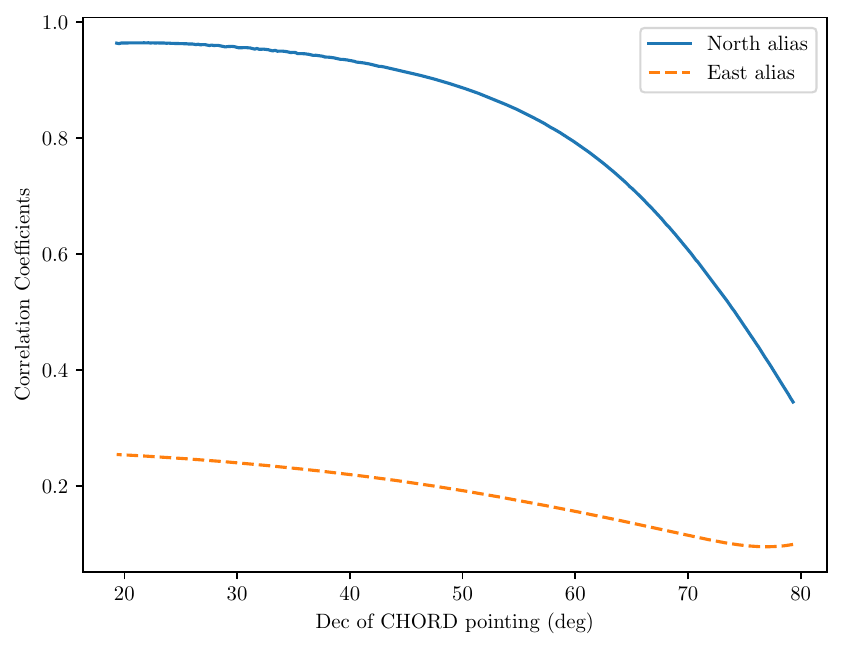}
\caption{The relationship between alias correlation coefficient and the direction that CHORD is pointing in. For this scenario, the source is located at CHORD's declination exactly.
The strength of the alias to the north of the true source position is strongest
near the equator and weakens toward the north pole.  A weaker trend exists for the
east-west aliases (which are more significantly attenuated by primary-beam-crossing effects).}
\label{fig:varydec}
\end{figure}

Figure \ref{fig:varydec} shows that alias disambiguation in the integrated case is easier at higher declinations, which is due to curvature. CHORD can point anywhere from a Dec of approximately +20 degrees to +80 degrees. The east alias correlation coefficient is reduced to below 0.3 by the primary beam crossing effect regardless of declination, and it is therefore unlikely to cause mislocalization. The north and south aliases can be disambiguated once curvature is important, but at low declinations, they are almost indistinguishable from the true source.

This relationship is key to optimizing the CHORD scan strategy so that the aliasing issue is reduced. Scanning at high declination decreases the solid angle scanned each day, but also decreases the correlation coefficient of aliases of detected sources and increases the sensitivity, since the exposure time (signal collected) per unit area is larger. A lower alias correlation coefficient decreases the probability of alias mislocalization.

We have seen that integrating over time provides a mechanism to reduce the alias correlation coefficients below 1. However, there exists a limiting case where an alias correlation coefficient approaches 1 because neither the primary beam crossing effect nor the curvature effect apply. This is bad because galaxies in this corner of parameter space will be subject to high alias mislocalization probabilities.

\begin{figure}[htb!]
\plotone{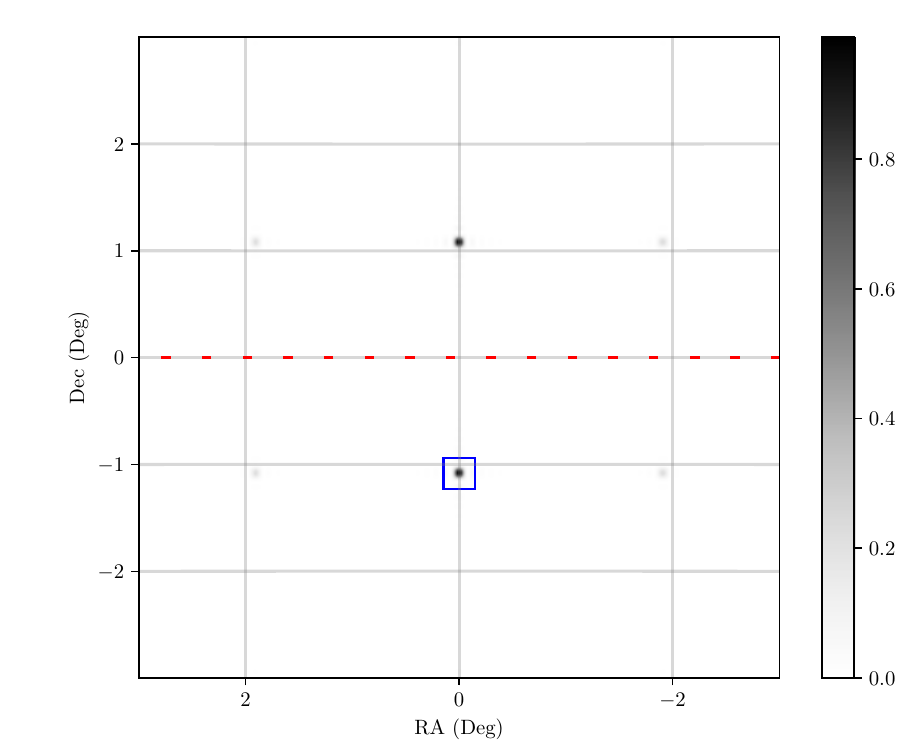}
\caption{This setup, with a hypothetical CHORD-like telescope that is capable of pointing at the equator and a source placed at exactly half the aliasing distance to the south, features an alias with a correlation coefficient of 1, even in the integrated case. The dashed line indicates CHORD's declination, and the box $\Box$ indicates the position of the source.}
\label{fig:disasterscenario}
\end{figure}

To produce this scenario, CHORD must be pointed at the equator in order to enter the flat sky regime. Then, the source must be placed a distance below CHORD such that its northern alias is the same distance above CHORD. In this scenario, the alias's and the source's primary beam response will be identical since their path through the primary beam is the same except reflected across the equator. Figure \ref{fig:disasterscenario} illustrates the situation.

Figure \ref{fig:disasterscenario} shows that an alias with a correlation coefficient of 1 is possible in the integrated case. CHORD will never actually point at the equator, but when it points at low declinations, then galaxies which are about this angular distance from CHORD's pointing declination will come with aliases that are nearly indistinguishable. This is the limit of using information inferred from pointing CHORD in a single direction. If we combine data from pointing CHORD in multiple directions, then this degeneracy can also be broken.

\section{Offset Pointing Observing}

CHORD's pointing can be changed in the elevation (Dec) direction but not in the azimuth (hour angle) direction; it always points along the meridian. If it is re-pointed by a very small amount multiple times, this is called dithering. A more likely scan strategy will be re-pointing by about the primary beam FWHM (2-2.5 deg) in order to observe multiple adjacent strips of sky. This choice of scan strategy has beneficial consequences for the galaxy search. The process of re-pointing must be done manually for all 512 dishes, and it is estimated that this will take about a week. Furthermore, changing direction might be counterproductive for some of CHORD's other science cases. This option is costly, but it has enormous power for alias disambiguation.

Re-pointing enters the matched filter equation as another dimension of the template and visibility arrays. The arrays depend on CHORD's pointing $\boldsymbol n_p$, which is equivalent to its declination. The dot product between the template and the visibility data can be extended to also sum over the $n_\theta$ re-pointing angles $\theta$. The correlation coefficient equation can be generalized to
\begin{multline}\label{eqn:ditheringmain}
    R(\boldsymbol n, \boldsymbol n_s) = \frac{1}{n_\tau n_\theta m_1^2 m_2^2}
    \frac{1}{\sqrt{\sum_{\tau,\theta} B\left(\abs{\mathcal R(\omega \tau)\boldsymbol n - \boldsymbol n_p(\theta)}\right)^4}\sqrt{\sum_{\tau,\theta} B\left(\abs{\mathcal R(\omega \tau)\boldsymbol n_s - \boldsymbol n_p(\theta)}\right)^4}}\\
    \times 
    \sum_{\tau, \theta} B\left(\abs{\mathcal R(\omega \tau)\boldsymbol n - \boldsymbol n_p(\theta)}\right)^2 B\left(\abs{\mathcal R(\omega \tau)\boldsymbol n_s - \boldsymbol n_p(\theta)}\right)^2 \\
    \times
    \frac{\sin(\pi m_1 L_1 /\lambda \ \boldsymbol e_1\cdot R(\omega \tau)(-\boldsymbol n+\boldsymbol n_s))^2}{\sin(\pi L_1 / \lambda \ \boldsymbol e_1 \cdot R(\omega \tau)(-\boldsymbol n+\boldsymbol n_s))^2}\frac{\sin(\pi m_2 L_2 / \lambda \ \boldsymbol e_2\cdot R(\omega \tau)(-\boldsymbol n+\boldsymbol n_s))^2}{\sin(\pi L_2 / \lambda \ \boldsymbol e_2 \cdot R(\omega \tau)(-\boldsymbol n+\boldsymbol n_s))^2}.
\end{multline}

\begin{figure}[htb!]
\plotone{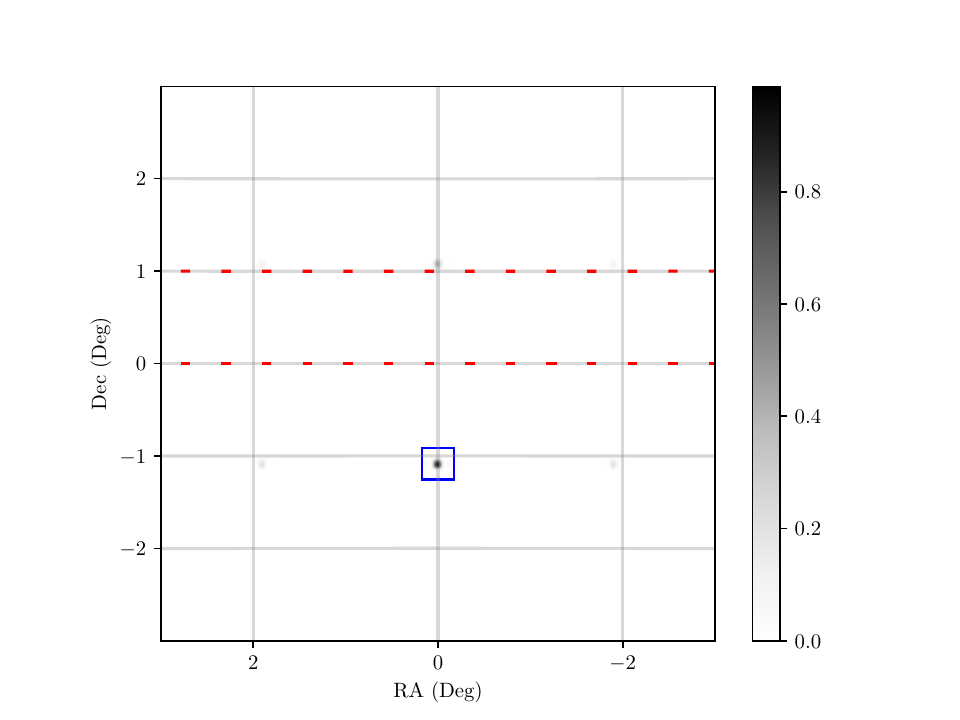}
\caption{A recreation of Figure \ref{fig:disasterscenario} but showing the correlation coefficient with a single offset upward by 1 degree.
In this scenario, CHORD is pointed at the equator and the source is in the worst-case aliasing position, resulting
in a perfect alias without the offset.
The dashed lines show the two declinations which CHORD is pointed at: 0 and 1 degrees.  With this 1-degree offset, the alias correlation coefficient is significantly reduced.}
\label{fig:dithering}
\end{figure}

The correlation coefficient is computed by summing over multiple complete Earth rotations, each with CHORD pointed at a different direction. If the plot of the worst-case scenario (Figure \ref{fig:disasterscenario}) is remade with two adjacent strips, then the previous alias with correlation coefficient 1 would have a correlation coefficient less than 1 (Figure \ref{fig:dithering}). This is because, as can be seen in Figure \ref{fig:deltatheta}, the alias correlation coefficients change depending on the relative positions of the source and CHORD. A near-degenerate alias in the case of one relative position will not be nearly degenerate when CHORD is pointed somewhere else.

\begin{figure}[htb!]
\plotone{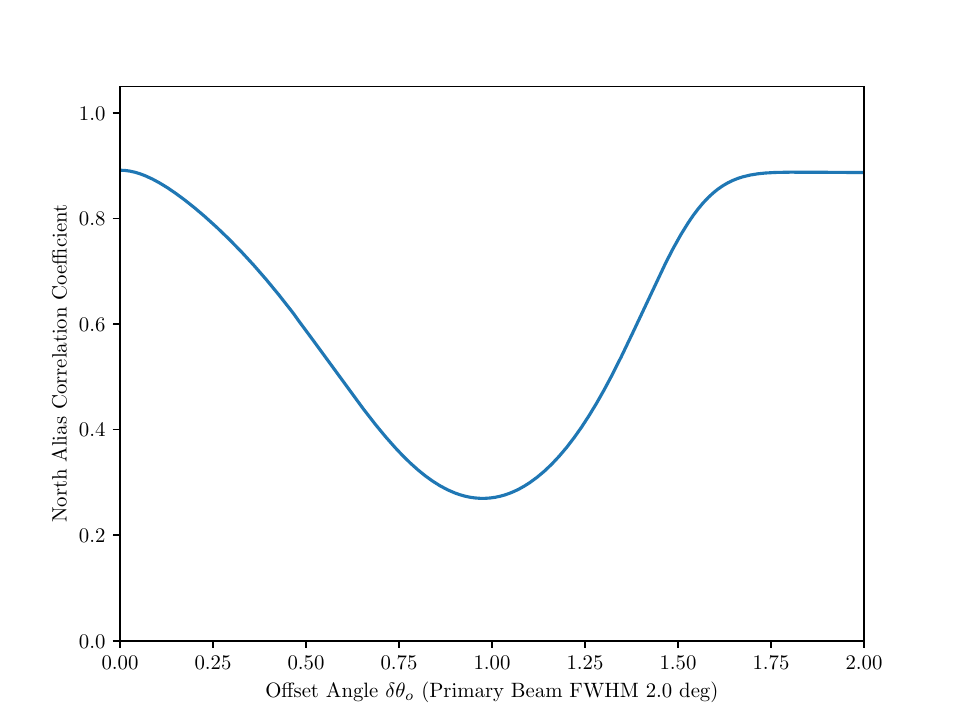}
\caption{How the northern alias's correlation coefficient changes depending on how much CHORD's pointing is offset between strips (i.e. how much the two dotted lines are separated by in Figure \ref{fig:dithering}). CHORD and the source are both located at Zenith.
We are considering only one additional strip; a single small dither results in measuring nearly the same signal and provides almost no additional information.  On the other hand, a very large offset means that both the true source position and the alias position are attenuated by the primary beam, so it is again impossible to distinguish between them.
The minimum correlation coefficient occurs at an offset angle of $\delta\theta_o = 2$ degrees.}
\label{fig:optimaldangle}
\end{figure}

The adjacent strip scan strategy introduces a new controllable $\delta \theta_o$, which is the angular offset distance that CHORD is re-pointed between strips. If $\delta \theta_o$ is too low, then the second strip is almost a repeated measurement and no new information is added. If $\delta \theta_o$ is too high, then CHORD at its new pointing will not pick up any signal from the source and cannot help. Therefore, given a single offset, there must be some optimal $\delta\theta_o$ to improve alias disambiguation. Figure \ref{fig:optimaldangle} demonstrates that for the best alias disambiguation, CHORD should be re-pointed to the locations of the aliases. This location depends on the position of the source, so Figure \ref{fig:optimaldangle} does not have general applicability. In the common case where the source is close to where CHORD is pointing, then the optimal angle becomes about 2 degrees. From an alias disambiguation perspective, ideally, the entire sky would be covered in 2 degree offset strips.

\begin{figure}[htb!]
\centering
\includegraphics[height=0.4\textwidth]{"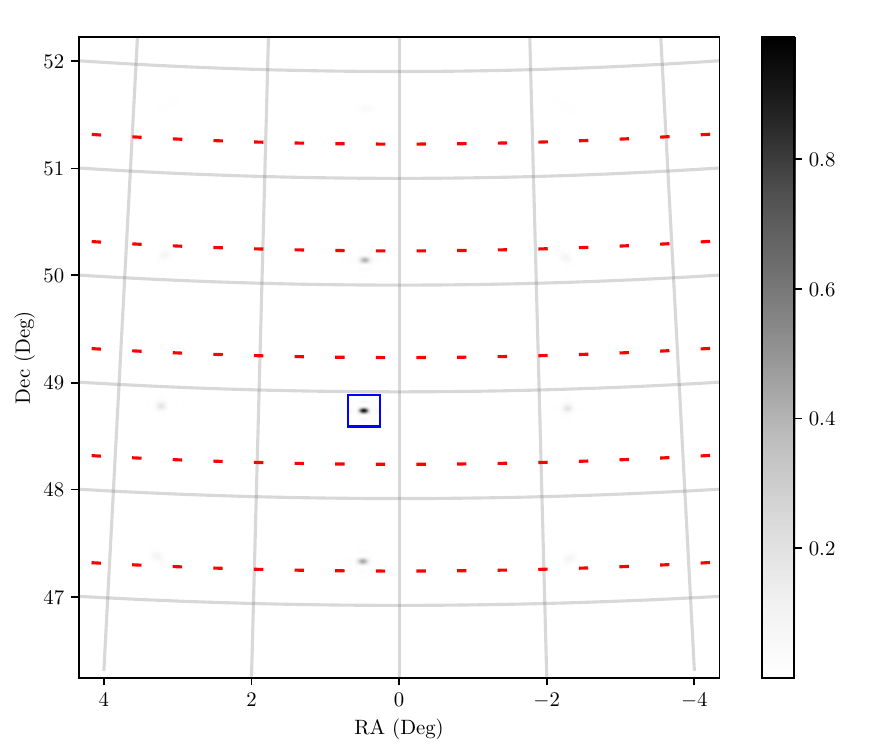"}
\includegraphics[height=0.4\textwidth]{"instantaneous_fig.pdf"}
\caption{\emph{Left:} The correlation coefficients given by Equation \ref{eqn:ditheringmain} for a typical declination and source position. CHORD is offset four times, and the adjacent pointings are indicated with dashed lines. The position of the true source is indicated by the box $\Box$. \emph{Right:} a copy of Figure \ref{fig:instantaneous}, the instantaneous case, for comparison.  A grid of perfect aliases exist in that case.  With time integration and adjacent-strip scanning, the aliases are significantly reduced.}
\label{fig:ditheringcomparison}
\end{figure}

Figure \ref{fig:ditheringcomparison} is a comparison between the instantaneous case (Figure \ref{fig:instantaneous}) and the same scenario but with time integration and adjacent-strip scanning in order to show how both sources of new information allow the true source to be easily picked out.

If there was no noise in the real data, then aliasing would not be a problem. After the matched filter is run, picking the highest value would always be the position of the real source. This is not the case when the data includes random noise. In the next section, we will translate the correlation coefficients that we have found into probabilities of alias mislocalization.

\section{Probabilistic Interpretation of Alias Correlation Coefficient}

After the matched filter is run, pixels with large values are selected as locations of galaxies. After, surrounding pixels as well as the pixels in alias locations must be labelled as belonging to the identified galaxy so that they are not searched again. This is the final step of galaxy detection and it can fail in two different ways. First, it can fail if the galaxy's brightness does not meet a signal to noise threshold. Second, it can fail if one of the aliases appears brighter than the pixel corresponding to the true source. In the latter case, the galaxy's reported position would be wrong and its spectrum would have the wrong scaling, making the catalogue entry harmful to most science cases. This result (mislocalization) can occur due to a random noise draw boosting a pixel where an alias happens to be located.

In this section, we want to quantify the probability of this outcome. To do this, we will assume that we are able to correctly group all the pixels in a synthesized peak together such that the true source and alias peaks are evaluated as a single number. For this analysis, we will assume that number is the reading of the highest value pixel in the group. Then, from the correlation coefficient map, we can determine the probability distribution of the matched filter values of all aliases.

Only the four closest aliases are going to be considered for this analysis. In order for alias mislocalization to occur, you must believe that the detected source brightness is larger than the source's real brightness by the inverse of the primary beam evaluated at the source position. Any north or south aliases which are far away from CHORD's pointing could be immediately ruled out simply because a source would have to be too bright to produce a response that far away. Also, it can be determined with the correlation coefficient maps that the east and west aliases have low significance aside from the first. The four points of the most nearby aliases will be treated as the only alternative mistaken locations for the source, with probabilities determined by their correlation coefficients.

\begin{figure}[htb!]
\centerline{\includegraphics[width=10cm]{"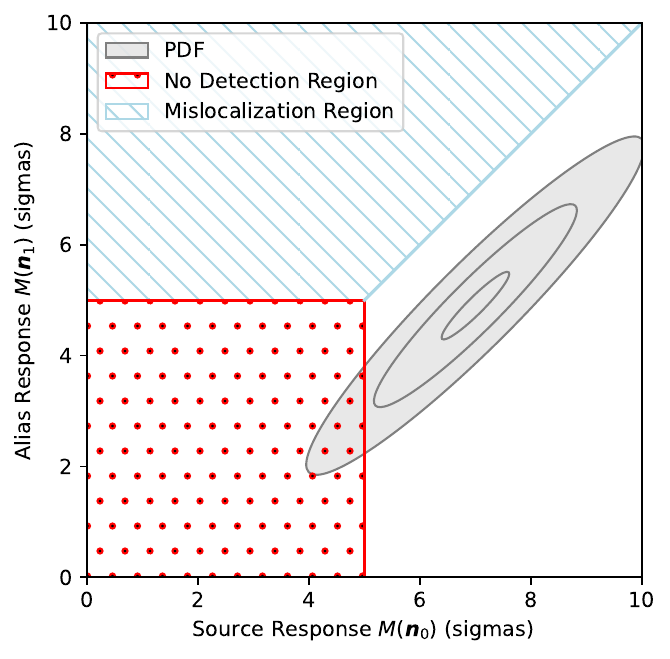"}}
\caption{An illustration of the 1, 3, and 5 $\sigma$ contours of the 2-dimensional PDF of the matched filter value of a source and an alias with important regions highlighted. The source has 7$\sigma$ in brightness, and the alias's correlation coefficient is 0.7. A detection threshold is shown at 5$\sigma$, and the region above the $M(u_0) = M(u_1)$ line is the case where the alias is brighter than the true source, i.e. the alias mislocalization region. A higher correlation coefficient shifts the center of the PDF closer to this line, but it also makes the contours thinner. For each detected galaxy, a value is drawn from a similar PDF, and unlucky draws will cause the final catalogue to lack completeness or reliability.}
\label{fig:probabilityregions}
\end{figure}

Because the matched filter is a linear operation on normally distributed data, the probability distribution function (PDF) of the matched filter value for each pixel is normal. Its definition is chosen so that the standard deviation of the PDF is 1. Since the matched filter values of the peak pixels are correlated with that of the source pixel, the 5-dimensional PDF of the matched filter value for the source and the four aliases is a correlated normal distribution. The probability that a random draw would fall below a detection threshold or would result in one of the alias matched filter values being larger than the source's can be computed numerically. A 2-dimensional (only looking at the true source and one alias) example is illustrated in Figure \ref{fig:probabilityregions}.

\begin{figure}[htb!]
\centerline{\includegraphics[width=10cm]{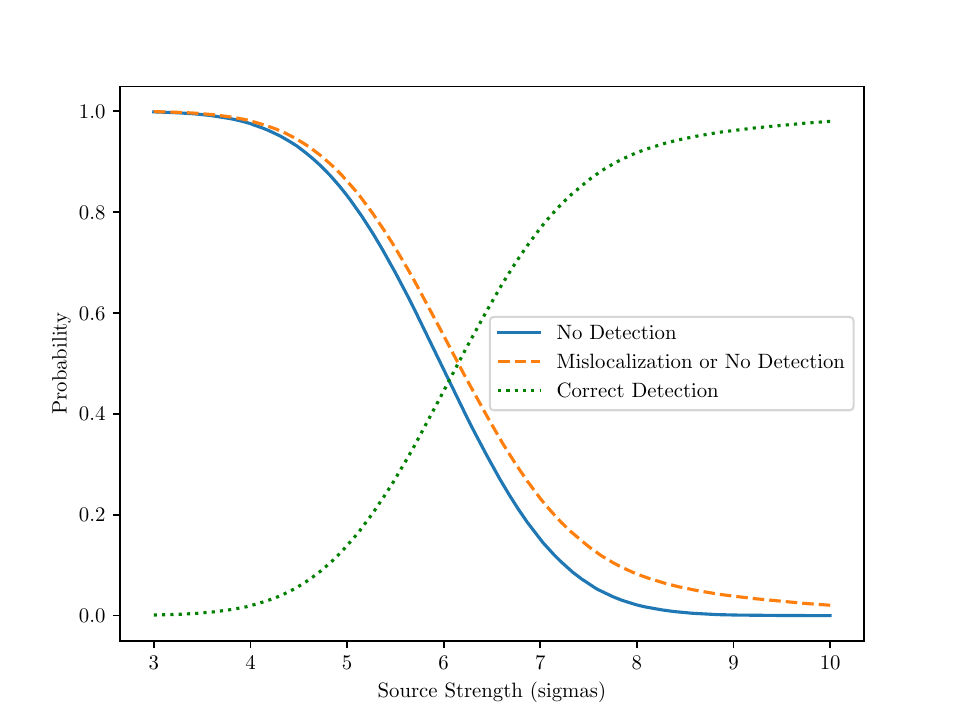}}
\caption{A prediction of the fraction of missed and mislocalized galaxies depending on the strength of the source. CHORD is pointed at zenith and the source is at the same declination. The detection threshold is 6$\sigma$.}
\label{fig:varysourcestrength}
\end{figure}

An example is illustrated in Figure \ref{fig:varysourcestrength}. It fixes the alias correlation coefficients and the detection threshold and varies the source strength. Then, it samples a similar PDF as shown in \ref{fig:probabilityregions}. The extremes of this plot are predictable. As the source strength approaches 0$\sigma$, it will not overcome noise and the detection probability is 0. As the source strength approaches infinity, the probability of the source dropping below the noise level approaches 0, and the probability of alias mislocalization is also low as long as the correlation coefficient is less than 1. Even for very high correlation coefficients, the fraction that is uncorrelated is unlikely to make up for the drop in response value for high source strengths because the mean of the Gaussian is far away from the mislocalization region.

\begin{figure}[htb!]
\centerline{\includegraphics[width=14cm]{"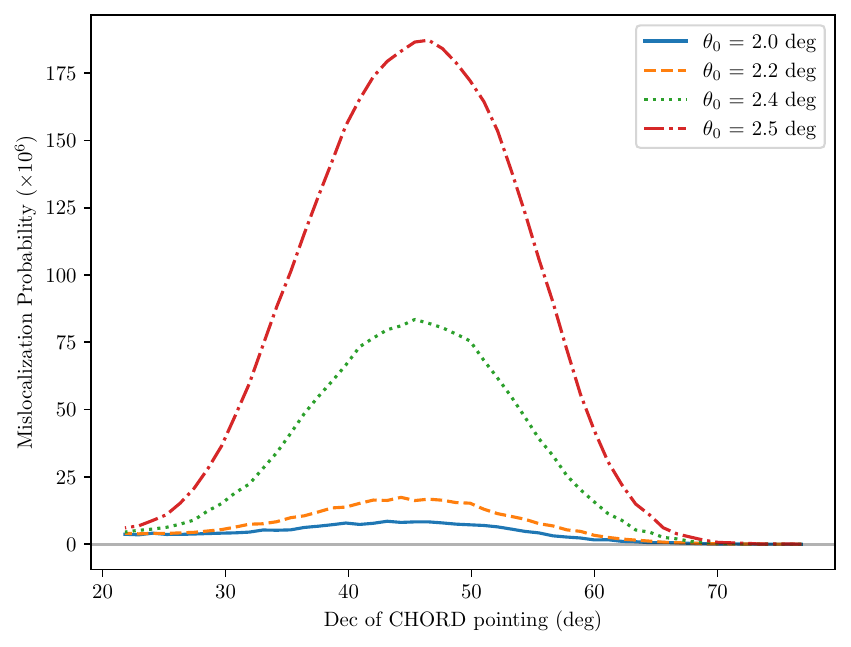"}}
\caption{A prediction of the probability of mislocalizing galaxies depending on where we point CHORD and using the adjacent strips scan strategy. For the simulation, CHORD is scanning along three strips separated by $\theta_0$ degrees, and the central strip is at the same declination as the source. The detection threshold is 6$\sigma$ and the source strength is 7$\sigma$. The jaggedness is caused by Monte-Carlo sampling.}
\label{fig:varydecprob}
\end{figure}

With this method, previous results that were stated in terms of the correlation coefficient can be restated in terms of the probability of the three outcomes: a correct detection, a non-detection, or mislocalization. Figure \ref{fig:varydecprob} shows the fraction of galaxies that will be mislocalized as a function of CHORD's declination if the adjacent strips scan strategy is employed. As seen in Figure \ref{fig:varydec}, at low declinations, the correlation coefficients of the north and south aliases are high if the single strip scan strategy is used. This would cause a high chance of mislocalization at low declinations. However, foreshortening effectively decreases the length of the north-south grid, moving the aliases closer to one of the adjacent strips, which decreases its correlation coefficient. This suggests that, although mislocalization is more difficult to prevent at low declinations in the single strip case, in the adjacent-strip-scanning case the positions of the strips relative to the source's aliases becomes a more important factor.

\begin{figure}[htb!]
\centerline{\includegraphics[width=14cm]{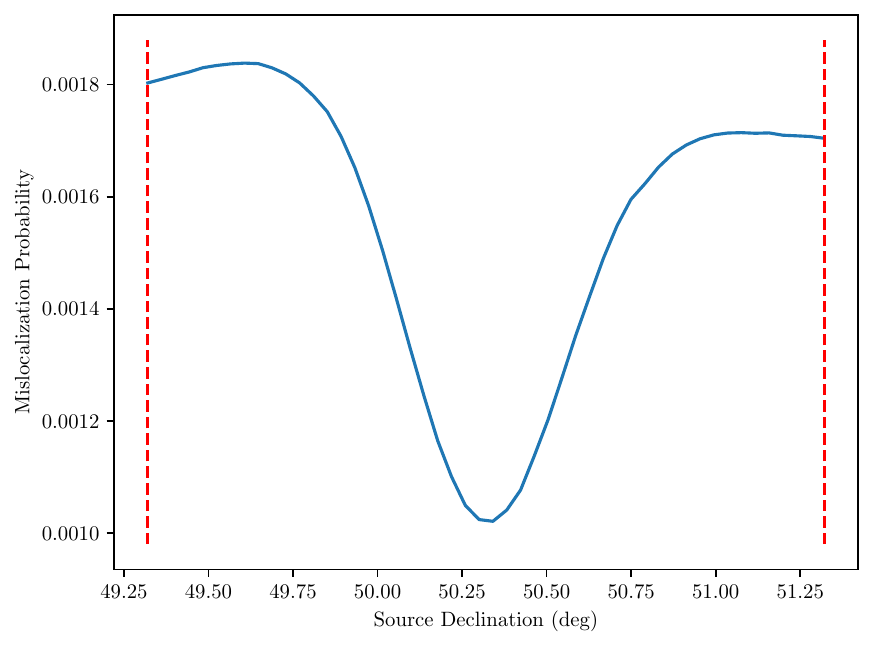}}
\caption{CHORD scans at two adjacent strips separated by 2 degrees indicated by the dashed lines. Sources in between the two pointing declinations will produce aliases with correlation coefficients that are diminished if the aliases are nearby to one of the pointing declinations. This results in a lower chance of mislocalization for sources in the middle of the two strips which have both north and south aliases suppressed.}
\label{fig:varydecbetweendithers}
\end{figure}

The variable mislocalization probability relative to CHORD's pointings is especially important in the eventuality where CHORD will only be repointed to a nearby position once. This situation is presented in Figure \ref{fig:varydecbetweendithers}. The second strip can only help reduce the alias correlation coefficient if the alias is within its primary beam FWHM scale. The strips in this example are separated by an offset angle of 2 degrees. This creates a situation where sources detected near the middle of the two strips have both of their correlation coefficients diminished, but sources near either pointing will have one difficult to distinguish alias. The resulting catalogue would have a strong declination dependence of mislocalized entries.

These probabilities allow us to predict what percentage of the final catalogue will be problematic. The catalogue can record cases where alias mislocalization is likely so that the end user can make an informed decision about which galaxies to consider. These predictions can also help inform where to point CHORD in order to minimize alias mislocalization.

\section{Conclusion}

A maximimally redundant array design exacerbates the spatial aliasing problem of interferometry. For the galaxy-survey science case where smooth spectral information cannot be leveraged to distinguish aliases, we find that there is a significant risk of misidentifying an aliased position as the true position of a galaxy. We investigate how a matched filter algorithm would be applied on the data of a transit redundant array telescope and how to predict the severity of the spatial aliases. The equation for the correlation coefficient shows that an instantaneous snapshot will produce fully ambiguous aliases, but that combining information from a transit over time, especially if the telescope can change orientation, will make it easier to choose the correct position for the source.

For the CHORD galaxy search, aliases will appear in a rectangular grid separated by about 2 degrees. With sufficient sampling of snapshots, only the four most nearby aliases are likely to cause mislocalization. If it is desired to offset CHORD to reduce the likelihood of mislocalization even more, CHORD should be offset by about two degrees for the most effect.

After the matched filter computation step, the quantity $M(\boldsymbol n;d)$ must be converted into a list of galaxies. Peaks in $M(\boldsymbol n;d)$ represent locations where point sources are likely to be located given visibilities $d$, so a looping argmax algorithm can be employed to return galaxy location candidates. When a source is identified, we must make sure that we do not pick nearby brightened pixels or the aliased locations associated with that galaxy in subsequent passes. This can be done by keeping track of which pixels are already associated with a galaxy, and then excluding those pixels in the future. Ignoring the spectral analysis, this is the process that would populate a galaxy catalogue. A full matched filter implementation will take advantage of both this spatial information as well as the spectral information of each source. Each entry would consist of a CHORD localisation, a spectrum (which is the maximum likelihood map), and the probabilities that the galaxy's location is actually a location that the algorithm decides is an alias.

The paper by Bij \etal\ (in prep.)  extrapolates the ALFALFA HI mass function and uses the current best measurements of the CHORD system sensitivity in order to estimate the number of galaxies that CHORD will detect in its full survey. The final count will depend on where CHORD is pointed and how often it is re-pointed. When CHORD is pointed farther north, on top of this reducing aliasing, the time it spends observing each unit area increases. This will increase sensitivity and cause lower brightness galaxies to be detected. On the other hand, re-pointing CHORD in order to observe more strips of sky, thus increasing the survey area, will lead to more galaxies being detected, albeit at lower depth.

An end-to-end simulation pipeline is under development, which can be used to quantify the effects that re-pointing will have on the final catalogue. Re-pointing CHORD will take several days which cannot be used for observing. We want to learn how beneficial re-pointing is for increasing counts and reducing the probability of mislocalization in exchange for this time cost.

\begin{acknowledgments}
    Research at Perimeter Institute is supported in part by the Government of Canada through the Department of Innovation, Science and Economic Development and by the Province of Ontario through the Ministry of Colleges, Universities, Research Excellence and Security. We acknowledge the support of the Natural Sciences and Engineering Research Council of Canada (NSERC), Discovery Grant RGPIN-2020-04254. KMS was supported by an NSERC Discovery Grant, by the Centre for the Universe at Perimeter Institute, and by the Daniel Family Foundation.
\end{acknowledgments}

\newpage

\bibliography{main}{}

@article{CHORDMainPaper,
  doi = {10.5281/ZENODO.3765414},
  url = {https://zenodo.org/record/3765414},
  author = {Vanderlinde, Keith and Liu, Adrian and Gaensler, Bryan and Bond, Dick and Hinshaw, Gary and Ng, Cherry and Chiang, Cynthia and Stairs, Ingrid and Brown, Jo-Anne and Sievers, Jonathan and Mena, Juan and Smith, Kendrick and Bandura, Kevin and Masui, Kiyoshi and Spekkens, Kristine and Belostotski, Leo and Dobbs, Matt and Turok, Neil and Boyle, Patrick and Rupen, Michael and Landecker, Tom and Pen, Ue-Li and Kaspi, Victoria},
  language = {en},
  title = {The Canadian Hydrogen Observatory and Radio-transient Detector (CHORD)},
  publisher = {Zenodo},
  year = {2019},
  copyright = {Creative Commons Attribution 4.0 International},
  journal = {Zenodo}
}

@ARTICLE{alfalfafull,
       author = {{Haynes}, Martha P. and {Giovanelli}, Riccardo and {Kent}, Brian R. and {Adams}, Elizabeth A.~K. and {Balonek}, Thomas J. and {Craig}, David W. and {Fertig}, Derek and {Finn}, Rose and {Giovanardi}, Carlo and {Hallenbeck}, Gregory and {Hess}, Kelley M. and {Hoffman}, G. Lyle and {Huang}, Shan and {Jones}, Michael G. and {Koopmann}, Rebecca A. and {Kornreich}, David A. and {Leisman}, Lukas and {Miller}, Jeffrey and {Moorman}, Crystal and {O'Connor}, Jessica and {O'Donoghue}, Aileen and {Papastergis}, Emmanouil and {Troischt}, Parker and {Stark}, David and {Xiao}, Li},
        title = "{The Arecibo Legacy Fast ALFA Survey: The ALFALFA Extragalactic H I Source Catalog}",
      journal = {Astrophysical Journal},
         year = 2018,
        month = jul,
       volume = {861},
       number = {1},
          eid = {49},
        pages = {49},
          doi = {10.3847/1538-4357/aac956},
archivePrefix = {arXiv},
       eprint = {1805.11499},
 primaryClass = {astro-ph.GA},
       adsurl = {https://ui.adsabs.harvard.edu/abs/2018ApJ...861...49H}
}

@article{Saintonge2007,
	doi = {10.1086/513515},
	url = {https://doi.org/10.1086%2F513515},
	year = 2007,
	month = {3},
	publisher = {American Astronomical Society},
	volume = {133},
	number = {5},
	pages = {2087--2096},
	author = {Am{\'{e}}lie Saintonge},
	title = {The Arecibo Legacy Fast {ALFA} Survey. {IV}. Strategies for Signal Identification and Survey Catalog Reliability},
	journal = {The Astronomical Journal}
}

@mastersthesis{hopkinsthesis,
    author = {Hans Hopkins},
    title = {Development of the CHORD Galaxy Search Strategy},
    school = {University of Waterloo},
    year = {2024},
    month = {9},
    url = {https://hdl.handle.net/10012/21089}
}

@ARTICLE{Giovanelli2015,
       author = {{Giovanelli}, Riccardo and {Haynes}, Martha P.},
        title = "{Extragalactic HI surveys}",
      journal = {\aapr},
         year = 2015,
        month = dec,
       volume = {24},
          eid = {1},
        pages = {1},
          doi = {10.1007/s00159-015-0085-3},
archivePrefix = {arXiv},
       eprint = {1510.04660},
 primaryClass = {astro-ph.GA},
       adsurl = {https://ui.adsabs.harvard.edu/abs/2015A&ARv..24....1G}
}

@misc{NRAOlectures,
title={NRAO Lectures Chapter 3: Radio Telescopes and Radiometers},
author={Condon, James J. and Ransom, Scott M},
year={2018},
howpublished={url: https://www.cv.nrao.edu/~sransom/web/Ch3.html}
}

@ARTICLE{HIPASS,
       author = {{Barnes}, D.~G. and {Staveley-Smith}, L. and {de Blok}, W.~J.~G. and {Oosterloo}, T. and {Stewart}, I.~M. and {Wright}, A.~E. and {Banks}, G.~D. and {Bhathal}, R. and {Boyce}, P.~J. and {Calabretta}, M.~R. and {Disney}, M.~J. and {Drinkwater}, M.~J. and {Ekers}, R.~D. and {Freeman}, K.~C. and {Gibson}, B.~K. and {Green}, A.~J. and {Haynes}, R.~F. and {te Lintel Hekkert}, P. and {Henning}, P.~A. and {Jerjen}, H. and {Juraszek}, S. and {Kesteven}, M.~J. and {Kilborn}, V.~A. and {Knezek}, P.~M. and {Koribalski}, B. and {Kraan-Korteweg}, R.~C. and {Malin}, D.~F. and {Marquarding}, M. and {Minchin}, R.~F. and {Mould}, J.~R. and {Price}, R.~M. and {Putman}, M.~E. and {Ryder}, S.~D. and {Sadler}, E.~M. and {Schr{\"o}der}, A. and {Stootman}, F. and {Webster}, R.~L. and {Wilson}, W.~E. and {Ye}, T.},
        title = {The HI Parkes All Sky Survey: southern observations, calibration and robust imaging},
      journal = {Monthly Notices of the RAS},
         year = 2001,
        month = apr,
       volume = {322},
       number = {3},
        pages = {486-498},
          doi = {10.1046/j.1365-8711.2001.04102.x},
       adsurl = {https://ui.adsabs.harvard.edu/abs/2001MNRAS.322..486B}
}

@ARTICLE{MA2025,
       author = {{Ma}, Wenlin and {Guo}, Hong and {Xu}, Haojie and {Jones}, Michael G. and {Zhang}, Chuan-Peng and {Zhu}, Ming and {Wang}, Jing and {Wang}, Jie and {Jiang}, Peng},
        title = "{The H I mass function of the Local Universe: Combining measurements from HIPASS, ALFALFA, and FASHI}",
      journal = {\aap},
         year = 2025,
        month = mar,
       volume = {695},
          eid = {A241},
        pages = {A241},
          doi = {10.1051/0004-6361/202452976},
archivePrefix = {arXiv},
       eprint = {2411.09903},
 primaryClass = {astro-ph.GA},
       adsurl = {https://ui.adsabs.harvard.edu/abs/2025A&A...695A.241M}
}

@ARTICLE{WALLABY,
       author = {{Koribalski}, B{\"a}rbel S. and {Staveley-Smith}, L. and {Westmeier}, T. and {Serra}, P. and {Spekkens}, K. and {Wong}, O.~I. and {Lee-Waddell}, K. and {Lagos}, C.~D.~P. and {Obreschkow}, D. and {Ryan-Weber}, E.~V. and {Zwaan}, M. and {Kilborn}, V. and {Bekiaris}, G. and {Bekki}, K. and {Bigiel}, F. and {Boselli}, A. and {Bosma}, A. and {Catinella}, B. and {Chauhan}, G. and {Cluver}, M.~E. and {Colless}, M. and {Courtois}, H.~M. and {Crain}, R.~A. and {de Blok}, W.~J.~G. and {D{\'e}nes}, H. and {Duffy}, A.~R. and {Elagali}, A. and {Fluke}, C.~J. and {For}, B. -Q. and {Heald}, G. and {Henning}, P.~A. and {Hess}, K.~M. and {Holwerda}, B.~W. and {Howlett}, C. and {Jarrett}, T. and {Jones}, D.~H. and {Jones}, M.~G. and {J{\'o}zsa}, G.~I.~G. and {Jurek}, R. and {J{\"u}tte}, E. and {Kamphuis}, P. and {Karachentsev}, I. and {Kerp}, J. and {Kleiner}, D. and {Kraan-Korteweg}, R.~C. and {L{\'o}pez-S{\'a}nchez}, {\'A}. R. and {Madrid}, J. and {Meyer}, M. and {Mould}, J. and {Murugeshan}, C. and {Norris}, R.~P. and {Oh}, S. -H. and {Oosterloo}, T.~A. and {Popping}, A. and {Putman}, M. and {Reynolds}, T.~N. and {Rhee}, J. and {Robotham}, A.~S.~G. and {Ryder}, S. and {Schr{\"o}der}, A.~C. and {Shao}, Li and {Stevens}, A.~R.~H. and {Taylor}, E.~N. and {van{\^A} der Hulst}, J.~M. and {Verdes-Montenegro}, L. and {Wakker}, B.~P. and {Wang}, J. and {Whiting}, M. and {Winkel}, B. and {Wolf}, C.},
        title = "{WALLABY - an SKA Pathfinder HI survey}",
      journal = {Astrophysics and Space Science},
         year = 2020,
        month = jul,
       volume = {365},
       number = {7},
          eid = {118},
        pages = {118},
          doi = {10.1007/s10509-020-03831-4},
archivePrefix = {arXiv},
       eprint = {2002.07311},
 primaryClass = {astro-ph.GA},
       adsurl = {https://ui.adsabs.harvard.edu/abs/2020Ap&SS.365..118K}
}

@ARTICLE{2018_HIMF,
       author = {{Jones}, Michael G. and {Haynes}, Martha P. and {Giovanelli}, Riccardo and {Moorman}, Crystal},
        title = "{The ALFALFA H I mass function: a dichotomy in the low-mass slope and a locally suppressed `knee' mass}",
      journal = {Monthly Notices of the RAS},
         year = 2018,
        month = jun,
       volume = {477},
       number = {1},
        pages = {2-17},
          doi = {10.1093/mnras/sty521},
archivePrefix = {arXiv},
       eprint = {1802.00053},
 primaryClass = {astro-ph.GA},
       adsurl = {https://ui.adsabs.harvard.edu/abs/2018MNRAS.477....2J}
}

@ARTICLE{HItoStellarMass,
       author = {{Huang}, Shan and {Haynes}, Martha P. and {Giovanelli}, Riccardo and {Brinchmann}, Jarle},
        title = "{The Arecibo Legacy Fast ALFA Survey: The Galaxy Population Detected by ALFALFA}",
      journal = {Astrophysical Journal},
         year = 2012,
        month = sep,
       volume = {756},
       number = {2},
          eid = {113},
        pages = {113},
          doi = {10.1088/0004-637X/756/2/113},
archivePrefix = {arXiv},
       eprint = {1207.0523},
 primaryClass = {astro-ph.CO},
       adsurl = {https://ui.adsabs.harvard.edu/abs/2012ApJ...756..113H}
}

@article{Lin_2023,
doi = {10.3847/1538-4357/accea2},
url = {https://doi.org/10.3847/1538-4357/accea2},
year = {2023},
month = {oct},
publisher = {The American Astronomical Society},
volume = {956},
number = {2},
pages = {148},
author = {Lin, Xuchen and Wang, Jing and Kilborn, Virginia and Peng, Eric W. and Cortese, Luca and Boselli, Alessandro and Liang, Ze-Zhong and Lee, Bumhyun and Yang, Dong and Catinella, Barbara and Deg, N. and Dénes, H. and Elagali, Ahmed and Kamphuis, P. and Koribalski, B. S. and Lee-Waddell, K. and Rhee, Jonghwan and Shao, Li and Spekkens, Kristine and Staveley-Smith, Lister and Westmeier, T. and Wong, O. Ivy and Bekki, Kenji and Bosma, Albert and Du, Min and Ho, Luis C. and Madrid, Juan P. and Verdes-Montenegro, Lourdes and Wang, Huiyuan and Wang, Shun},
title = {FAST-ASKAP Synergy: Quantifying Coexistent Tidal and Ram Pressure Strippings in the NGC 4636 Group},
journal = {The Astrophysical Journal}
}

@misc{du2024opticallydarkgalaxiesdecals,
      title={Almost Optically Dark Galaxies in DECaLS (I): Detection, Optical Properties and Possible Origins}, 
      author={Lin Du and Wei Du and Cheng Cheng and Ming Zhu and Haiyang Yu and Hong Wu},
      year={2024},
      eprint={2403.12130},
      archivePrefix={arXiv},
      primaryClass={astro-ph.GA},
      url={https://arxiv.org/abs/2403.12130}, 
}

@article{Franco_2025,
   title={Measuring the matter fluctuations in the Local Universe with the ALFALFA catalogue},
   volume={537},
   ISSN={1365-2966},
   url={http://dx.doi.org/10.1093/mnras/staf088},
   DOI={10.1093/mnras/staf088},
   number={2},
   journal={Monthly Notices of the Royal Astronomical Society},
   publisher={Oxford University Press (OUP)},
   author={Franco, Camila and Oliveira, Jezebel and Lopes, Maria and Avila, Felipe and Bernui, Armando},
   year={2025},
   month=jan, pages={897–908} }

@unpublished{Deg,
title= {Plenary Talk},
author = {Deg, Nathan},
year = {2025},
note= {GEESE-ON Workshop}
}

@article{chimeOverview,
   title={An Overview of CHIME, the Canadian Hydrogen Intensity Mapping Experiment},
   volume={261},
   ISSN={1538-4365},
   url={http://dx.doi.org/10.3847/1538-4365/ac6fd9},
   DOI={10.3847/1538-4365/ac6fd9},
   number={2},
   journal={The Astrophysical Journal Supplement Series},
   publisher={American Astronomical Society},
   author={Amiri, Mandana and Bandura, Kevin and Boskovic, Anja and Chen, Tianyue and Cliche, Jean-François and Deng, Meiling and Denman, Nolan and Dobbs, Matt and Fandino, Mateus and Foreman, Simon and Halpern, Mark and Hanna, David and Hill, Alex S. and Hinshaw, Gary and Höfer, Carolin and Kania, Joseph and Klages, Peter and Landecker, T. L. and MacEachern, Joshua and Masui, Kiyoshi and Mena-Parra, Juan and Milutinovic, Nikola and Mirhosseini, Arash and Newburgh, Laura and Nitsche, Rick and Ordog, Anna and Pen, Ue-Li and Pinsonneault-Marotte, Tristan and Polzin, Ava and Reda, Alex and Renard, Andre and Shaw, J. Richard and Siegel, Seth R. and Singh, Saurabh and Smegal, Rick and Tretyakov, Ian and Van Gassen, Kwinten and Vanderlinde, Keith and Wang, Haochen and Wiebe, Donald V. and Willis, James S. and Wulf, Dallas},
   year={2022},
   month=jul, pages={29} }

@article{Obuljen_2019,
   title={The H i content of dark matter haloes at z ≈ 0 from ALFALFA},
   volume={486},
   ISSN={1365-2966},
   url={http://dx.doi.org/10.1093/mnras/stz1118},
   DOI={10.1093/mnras/stz1118},
   number={4},
   journal={Monthly Notices of the Royal Astronomical Society},
   publisher={Oxford University Press (OUP)},
   author={Obuljen, Andrej and Alonso, David and Villaescusa-Navarro, Francisco and Yoon, Ilsang and Jones, Michael},
   year={2019},
   month=apr, pages={5124–5138} }

@ARTICLE{NVSS,
       author = {{Condon}, J.~J. and {Cotton}, W.~D. and {Greisen}, E.~W. and {Yin}, Q.~F. and {Perley}, R.~A. and {Taylor}, G.~B. and {Broderick}, J.~J.},
        title = "{The NRAO VLA Sky Survey}",
      journal = {\aj},
         year = 1998,
        month = may,
       volume = {115},
       number = {5},
        pages = {1693-1716},
          doi = {10.1086/300337},
       adsurl = {https://ui.adsabs.harvard.edu/abs/1998AJ....115.1693C}
}

@ARTICLE{FIRST,
       author = {{Becker}, Robert H. and {White}, Richard L. and {Helfand}, David J.},
        title = "{The FIRST Survey: Faint Images of the Radio Sky at Twenty Centimeters}",
      journal = {\apj},
         year = 1995,
        month = sep,
       volume = {450},
        pages = {559},
          doi = {10.1086/176166},
       adsurl = {https://ui.adsabs.harvard.edu/abs/1995ApJ...450..559B}
}

@ARTICLE{WallabyPilotPhase2,
       author = {{Murugeshan}, C. and {Deg}, N. and {Westmeier}, T. and {Shen}, A.~X. and {For}, B.-Q. and {Spekkens}, K. and {Wong}, O.~I. and {Staveley-Smith}, L. and {Catinella}, B. and {Lee-Waddell}, K. and {D{\'e}nes}, H. and {Rhee}, J. and {Cortese}, L. and {Goliath}, S. and {Halloran}, R. and {van der Hulst}, J.~M. and {Kamphuis}, P. and {Koribalski}, B.~S. and {Kraan-Korteweg}, R.~C. and {Lelli}, F. and {Venkataraman}, P. and {Verdes-Montenegro}, L. and {Yu}, N.},
        title = "{WALLABY Pilot Survey: Public data release of {\ensuremath{\sim}} 1800 H I sources and high-resolution cut-outs from Pilot Survey Phase 2}",
      journal = {\pasa},
         year = 2024,
        month = nov,
       volume = {41},
          eid = {e088},
        pages = {e088},
          doi = {10.1017/pasa.2024.91},
archivePrefix = {arXiv},
       eprint = {2409.13130},
 primaryClass = {astro-ph.GA},
       adsurl = {https://ui.adsabs.harvard.edu/abs/2024PASA...41...88M}
}

@ARTICLE{tully-fisher-alfalfa,
       author = {{Papastergis}, E. and {Adams}, E.~A.~K. and {van der Hulst}, J.~M.},
        title = "{An accurate measurement of the baryonic Tully-Fisher relation with heavily gas-dominated ALFALFA galaxies}",
      journal = {\aap},
         year = 2016,
        month = sep,
       volume = {593},
          eid = {A39},
        pages = {A39},
          doi = {10.1051/0004-6361/201628410},
archivePrefix = {arXiv},
       eprint = {1602.09087},
 primaryClass = {astro-ph.GA},
       adsurl = {https://ui.adsabs.harvard.edu/abs/2016A&A...593A..39P}
}

@INPROCEEDINGS{CHIMEFRBMain,
       author = {{Ng}, C. and {Vanderlinde}, K. and {Paradise}, A. and {Klages}, P. and {Masui}, K. and {Smith}, K. and {Bandura}, K. and {Boyle}, P.~J. and {Dobbs}, M. and {Kaspi}, V. and {Renard}, A. and {Shaw}, J.~R. and {Stairs}, I. and {Tretyakov}, I.},
        title = "{CHIME FRB: An application of FFT beamforming for a radio telescope}",
    booktitle = {XXXII International Union of Radio Science General Assembly \& Scientific Symposium (URSI GASS) 2017},
         year = 2017,
        month = aug,
          eid = {4},
        pages = {4},
          doi = {10.23919/URSIGASS.2017.8105318},
archivePrefix = {arXiv},
       eprint = {1702.04728},
 primaryClass = {astro-ph.IM},
       adsurl = {https://ui.adsabs.harvard.edu/abs/2017ursi.confE...4N}
}

@ARTICLE{BOSS,
       author = {{Bolton}, Adam S. and {Schlegel}, David J. and {Aubourg}, {\'E}ric and {Bailey}, Stephen and {Bhardwaj}, Vaishali and {Brownstein}, Joel R. and {Burles}, Scott and {Chen}, Yan-Mei and {Dawson}, Kyle and {Eisenstein}, Daniel J. and {Gunn}, James E. and {Knapp}, G.~R. and {Loomis}, Craig P. and {Lupton}, Robert H. and {Maraston}, Claudia and {Muna}, Demitri and {Myers}, Adam D. and {Olmstead}, Matthew D. and {Padmanabhan}, Nikhil and {P{\^a}ris}, Isabelle and {Percival}, Will J. and {Petitjean}, Patrick and {Rockosi}, Constance M. and {Ross}, Nicholas P. and {Schneider}, Donald P. and {Shu}, Yiping and {Strauss}, Michael A. and {Thomas}, Daniel and {Tremonti}, Christy A. and {Wake}, David A. and {Weaver}, Benjamin A. and {Wood-Vasey}, W. Michael},
        title = "{Spectral Classification and Redshift Measurement for the SDSS-III Baryon Oscillation Spectroscopic Survey}",
      journal = {\aj},
         year = 2012,
        month = nov,
       volume = {144},
       number = {5},
          eid = {144},
        pages = {144},
          doi = {10.1088/0004-6256/144/5/144},
archivePrefix = {arXiv},
       eprint = {1207.7326},
 primaryClass = {astro-ph.CO},
       adsurl = {https://ui.adsabs.harvard.edu/abs/2012AJ....144..144B}
}

@ARTICLE{redmapper,
       author = {{Rykoff}, E.~S. and {Rozo}, E. and {Busha}, M.~T. and {Cunha}, C.~E. and {Finoguenov}, A. and {Evrard}, A. and {Hao}, J. and {Koester}, B.~P. and {Leauthaud}, A. and {Nord}, B. and {Pierre}, M. and {Reddick}, R. and {Sadibekova}, T. and {Sheldon}, E.~S. and {Wechsler}, R.~H.},
        title = "{redMaPPer. I. Algorithm and SDSS DR8 Catalog}",
      journal = {\apj},
         year = 2014,
        month = apr,
       volume = {785},
       number = {2},
          eid = {104},
        pages = {104},
          doi = {10.1088/0004-637X/785/2/104},
archivePrefix = {arXiv},
       eprint = {1303.3562},
 primaryClass = {astro-ph.CO},
       adsurl = {https://ui.adsabs.harvard.edu/abs/2014ApJ...785..104R}
}

@ARTICLE{Melin,
       author = {{Melin}, J.-B. and {Bartlett}, J.~G. and {Delabrouille}, J.},
        title = "{Catalog extraction in SZ cluster surveys: a matched filter approach}",
      journal = {\aap},
         year = 2006,
        month = nov,
       volume = {459},
       number = {2},
        pages = {341-352},
          doi = {10.1051/0004-6361:20065034},
archivePrefix = {arXiv},
       eprint = {astro-ph/0602424},
 primaryClass = {astro-ph},
       adsurl = {https://ui.adsabs.harvard.edu/abs/2006A&A...459..341M}
}

@ARTICLE{Willman,
       author = {{Willman}, Beth and {Blanton}, Michael R. and {West}, Andrew A. and {Dalcanton}, Julianne J. and {Hogg}, David W. and {Schneider}, Donald P. and {Wherry}, Nicholas and {Yanny}, Brian and {Brinkmann}, Jon},
        title = "{A New Milky Way Companion: Unusual Globular Cluster or Extreme Dwarf Satellite?}",
      journal = {\aj},
         year = 2005,
        month = jun,
       volume = {129},
       number = {6},
        pages = {2692-2700},
          doi = {10.1086/430214},
archivePrefix = {arXiv},
       eprint = {astro-ph/0410416},
 primaryClass = {astro-ph},
       adsurl = {https://ui.adsabs.harvard.edu/abs/2005AJ....129.2692W}
}

@article{Doyle_2000,
   title={Observational Limits on Terrestrial‐sized Inner Planets around the CM Draconis System Using the Photometric Transit Method with a Matched‐Filter Algorithm},
   volume={535},
   ISSN={1538-4357},
   url={http://dx.doi.org/10.1086/308830},
   DOI={10.1086/308830},
   number={1},
   journal={The Astrophysical Journal},
   publisher={American Astronomical Society},
   author={Doyle, Laurance R. and Deeg, Hans J. and Kozhevnikov, Valerij P. and Oetiker, Brian and Martin, Eduardo L. and Blue, J. Ellen and Rottler, Lee and Stone, Remington P. S. and Ninkov, Zoran and Jenkins, Jon M. and Schneider, Jean and Dunham, Edward W. and Doyle, Moira F. and Paleologou, Efthimious},
   year={2000},
   month=may, pages={338–349} }

@ARTICLE{ligo,
   author = {{Abbott}, B.~P. and {Abbott}, R. and {Abbott}, T.~D. and {Abernathy}, M.~R. and 
	{Acernese}, F. and {Ackley}, K. and {Adams}, C. and {Adams}, T. and 
	{Addesso}, P. and {Adhikari}, R.~X. and et al.},
    title = "{Observation of Gravitational Waves from a Binary Black Hole Merger}",
  journal = {Physical Review Letters},
archivePrefix = "arXiv",
   eprint = {1602.03837},
 primaryClass = "gr-qc",
     year = 2016,
    month = feb,
   volume = 116,
   number = 6,
      eid = {061102},
    pages = {061102},
      doi = {10.1103/PhysRevLett.116.061102},
   adsurl = {http://adsabs.harvard.edu/abs/2016PhRvL.116f1102A}
}

@article{Liu_2020,
   title={Data Analysis for Precision 21 cm Cosmology},
   volume={132},
   ISSN={1538-3873},
   url={http://dx.doi.org/10.1088/1538-3873/ab5bfd},
   DOI={10.1088/1538-3873/ab5bfd},
   number={1012},
   journal={Publications of the Astronomical Society of the Pacific},
   publisher={IOP Publishing},
   author={Liu, Adrian and Shaw, J. Richard},
   year={2020},
   month=apr, pages={062001} }

@PROCEEDINGS{Perley,
        title = "{Synthesis imaging in radio astronomy : a collection of lectures from the third NRAO synthesis imaging summer school}",
    booktitle = {Synthesis Imaging in Radio Astronomy},
         year = 1989,
       editor = {{Perley}, Richard A. and {Schwab}, Frederic R. and {Bridle}, Alan H.},
       series = {Astronomical Society of the Pacific Conference Series},
       volume = {6},
        month = jan,
       adsurl = {https://ui.adsabs.harvard.edu/abs/1989ASPC....6.....P}
}

@ARTICLE{FASHI,
       author = {{Zhang}, Chuan-Peng and {Zhu}, Ming and {Jiang}, Peng and {Cheng}, Cheng and {Wang}, Jing and {Wang}, Jie and {Xu}, Jin-Long and {Liu}, Xiao-Lan and {Yu}, Nai-Ping and {Qian}, Lei and {Yu}, Haiyang and {Ai}, Mei and {Jing}, Yingjie and {Xu}, Chen and {Liu}, Ziming and {Guan}, Xin and {Sun}, Chun and {Yang}, Qingliang and {Huang}, Menglin and {Hao}, Qiaoli and {FAST Collaboration}},
        title = "{The FAST all sky H I survey (FASHI): The first release of catalog}",
      journal = {Science China Physics, Mechanics, and Astronomy},
         year = 2024,
        month = jan,
       volume = {67},
       number = {1},
          eid = {219511},
        pages = {219511},
          doi = {10.1007/s11433-023-2219-7},
archivePrefix = {arXiv},
       eprint = {2312.06097},
 primaryClass = {astro-ph.GA},
       adsurl = {https://ui.adsabs.harvard.edu/abs/2024SCPMA..6719511Z}
}
\bibliographystyle{aasjournalv7}



\end{document}